\documentclass[%
 reprint,
superscriptaddress,
showpacs,preprintnumbers,
nofootinbib,
 amsmath,amssymb,aps,
prd,
floatfix,
]{revtex4-1}

\usepackage{lipsum} 

\usepackage{bm} 
\usepackage{physics}
\usepackage{comment}
\pdfoutput=1
\usepackage{amsmath,amsfonts,amsthm}
\usepackage{esdiff} 
\usepackage{booktabs}  
\usepackage{url}  
\usepackage[hidelinks]{hyperref}  
\usepackage{relsize}

\usepackage{cleveref}  
	\crefname{equation}{equation}{equations}
	\crefname{figure}{figure}{figures}	
	\crefname{table}{table}{tables}
\usepackage[caption=false]{subfig}

\usepackage[normalem]{ulem}

\usepackage[usenames,dvipsnames,svgnames,table]{xcolor}

\usepackage[export]{adjustbox}
\usepackage{graphicx}

\usepackage{stfloats} 
\usepackage{placeins} 

\usepackage[sc]{mathpazo} 
\usepackage[T1]{fontenc} 
\usepackage{microtype} 

\usepackage{booktabs} 
\usepackage{float} 
\usepackage{hyperref} 

\usepackage{lettrine} 
\usepackage{paralist} 

\usepackage{titlesec} 
\renewcommand\thesection{\Roman{section}} 
\renewcommand\thesubsection{\Alph{subsection}} 
\titleformat{\section}[block]{\large\scshape\centering\bfseries}{\thesection.}{1em}{} 

\titleformat{\subsection}[block]{\scshape\centering}{\thesubsection.}{1em}{} 

\DeclareCaptionFormat{myformat}{#1#2#3\hrulefill}
\begin{document}

\title{Leveraging neutrino flavor physics for supernova model differentiation} 

\author{Lily Newkirk}\thanks{Corresponding author}
\email{lnewkirk@nyit.edu}
\author{Eve Armstrong}
\email{Deceased}
\affiliation{Department of Physics, New York Institute of Technology, New York, NY 10023, USA}
\affiliation{Department of Astrophysics, American Museum of Natural History, New York, NY 10024, USA}
\author{A. Baha Balantekin}
\email{baha@physics.wisc.edu}
\affiliation{Department of Physics, University of Wisconsin: Madison, Madison, WI 53706, USA}
\author{Adam Burrows}
\email{burrows@astro.princeton.edu}
\affiliation{Department of Astrophysical Sciences, Princeton University, Princeton, NJ 08544, USA}
\author{Yennaly F. Isiano}
\email{yferna05@nyit.edu}
\affiliation{Department of Physics, New York Institute of Technology, New York, NY 10023, USA}
\author{Elizabeth K. Jones}
\affiliation{Department of Physics, Harvey Mudd College, Claremont, CA 91711}
\author{Caroline Laber-Smith}
\affiliation{Department of Physics, University of Wisconsin: Madison, Madison, WI 53706, USA}
\author{Amol V.\ Patwardhan}
\email{apatwa02@nyit.edu} 
\affiliation{Department of Physics, New York Institute of Technology, New York, NY 10023, USA}
\affiliation{School of Physics \& Astronomy, University of Minnesota, Minneapolis, MN 55455}
\author{Sarah Ranginwala}
\affiliation{Department of Physics, New York Institute of Technology, New York, NY 10023, USA}
\author{Hansen Torres}
\affiliation{Department of Physics, New York Institute of Technology, New York, NY 10023, USA}

\date{\today}


\begin{abstract}
{Neutrino flavor evolution is critical for understanding the physics of dense astrophysical regimes, including core-collapse supernovae (CCSN).  Powerful numerical integration codes exist for simulating these environments, yet a complete understanding of the inherent nonlinearity of collective neutrino flavor oscillations and how it fits within the overall framework of these simulations remains an open challenge.  For this reason, we continue developing statistical data assimilation (SDA) to infer solutions to the flavor field in a CCSN envelope, given simulated measurements far from the source.  SDA is an inference paradigm designed to optimize a model with sparse data.  Our model consists of neutrino beams emanating from a CCSN and coherently interacting with each other and with a background of other matter particles in one dimension $r$.  One model feature of high interest is the distribution of those matter particles as a function of radius $r$, or the "matter potential" $V(r)$ -- as it significantly dictates flavor evolution.  In this paper, we expand the model beyond previous incarnations, by replacing the monotonically-decaying analytic form for $V(r)$ we previously used with a more complex -- and more physically plausible -- set of profiles derived from a one-dimensional (spherically symmetric) hydrodynamics simulation of a CCSN explosion.  We ask whether the SDA procedure can use simulated flavor measurements at physically accessible locations (i.e. in vacuum) to determine the extent to which different matter density profiles through which the neutrinos propagate in the matter-dominated regime are compatible with these measurements.  Within the scope of our small-scale model, we find that the neutrino flavor measurements in the vacuum regime are able to discriminate between different matter profiles, and we discuss implications regarding a future galactic CCSN detection.}

\end{abstract}


\maketitle

\section{Introduction} \label{sec:intro}

This era of multi-messenger astrophysics offers an unprecedented vantage point on core-collapse supernovae (CCSN) and binary neutron star mergers.  Neutrino fluxes and densities are high in these events, and thus neutrino flavor evolution can significantly shape the physics therein~\cite{Fuller:1992eu,Qian:1993dg,Fuller:1993ry,Fuller:1995qy,Duan:2010af,Wu:2014kaa,Wu:2015glr,Sasaki:2017jry,Balantekin:2017bau,Xiong:2019nvw,Xiong:2020ntn}.  Neutrino flavor evolution can be shaped by neutrino interactions with other matter particles, as well as with other neutrinos, the latter resulting in a variety of collective phenomena in flavor space (e.g. \cite{Duan:2009cd,Duan:2010bg,Mirizzi:2015eza,Chakraborty:2016yeg,Tamborra:2020cul,Richers:2022zug} and references therein.)  Recent advances in astrophysical modelling as well as detector development call for a deeper understanding of i) how neutrino flavor evolution impacts energy, entropy, and lepton number transport at these sites, and ii) the expected signatures thereof in neutrino, gravitational wave, and electromagnetic observations.  

Powerful numerical integration codes exist for obtaining solutions to the flavor evolution problem in compact object environments (e.g.,Refs.~\cite{duan2008simulating,richers2019neutrino, Richers:2021nbx, Bhattacharyya:2020jpj, Zaizen:2021wwl, Martin:2019kgi, George:2022lwg}; see also \cite{Richers:2022bkd} for a recent review).  These codes, however, require making choices regarding the relative flavor content of the neutrinos at emission from the proto-neutron star, based on reasonable physical assumptions about the nuclear matter and neutrino decoupling. 
Traditionally, in many of these calculations (pre-2015), the decoupling of neutrinos from chemical and thermal equilibrium was approximated to be instantaneous at the surface of the proto-neutron star (represented by a sharp \lq\lq neutrino-sphere\rq\rq), and independent of neutrino flavor and energy. Thus, the neutrinos at decoupling were taken to be in definite flavor states, at a fixed radial distance across all flavors and energies. Recently, however, relaxing these assumptions has significantly shifted the paradigm of flavor evolution simulations.  Specifically: there emerges an emission-angle dependence in the initial flavor content, causing \lq\lq fast\rq\rq\ flavor oscillations (see Refs.~\cite{Tamborra:2020cul,Richers:2022zug} and references therein.)  Thus it is apparent that \textit{a priori} assumptions on initial conditions -- a requirement of forward integration codes -- are undesirable.

A second disadvantage of the forward integration codes is their formulation as an initial-value problem.  This framework, particularly in a steady-state formulation, may not be ideal for capturing direction-changing scattering of neutrinos off other neutrinos -- a critical component of flavor physics.  This \lq\lq halo effect\rq\rq\ \cite{cherry2012neutrino,Cherry:2013mv,Zaizen:2019ufj,Cherry:2019vkv,cirigliano2018collective} can result in a non-outward-propagating component of the neutrino flux that can significantly contribute to the forward-scattering potential experienced by the outgoing neutrinos, as a result of the large intersection angles between their trajectories.  Thus the problem formulation becomes -- not an initial-value problem with flavor content fully specified at the source -- but rather a two-point boundary-value problem, with flavor information propagating both out and in.  

In this paper, we instead adopt an inverse approach.  Inference is a means to optimize a model with measurements, where the measurements are assumed to arise from model dynamics.  Within this framework, we assume partial information at the bounds together with a model, and we seek model regimes with which the measurements can be reconciled.

The specific inference technique used is statistical data assimilation (SDA).  SDA was invented for numerical Earth-based weather prediction~\cite{kimura2002numerical,kalnay2003atmospheric,evensen2009data,betts2010practical,whartenby2013number,an2017estimating} for the case of sparse data.  It has gained traction in neurobiology~\cite{schiff2009kalman,toth2011dynamical,kostuk2012dynamical,hamilton2013real,meliza2014estimating,nogaret2016automatic,armstrong2020statistical}, for estimating cellular and synaptic properties given sparse neuronal electrical signals.  We have brought SDA into astrophysics, to explore solutions to small-scale flavor evolution models~\cite{armstrong2017optimization,armstrong2020inference,rrapaj2021inference,armstrong2021inference,armstrong2022inferenceA,armstrong2022inferenceB,laber2023inference}.

Here we build on previous work, in our representation of the matter potential $V(r)$ within the CCSN envelope.  The matter density profile near a supernova depends on various environmental factors such as the mass of the progenitor star, the time elapsed after the initial core bounce, the details of the neutrino emission, the degree of symmetry, and the strength of the magnetic fields. This large degree of intrinsic variation in $V(r)$ across different CCSNe, or even at different times in the same CCSN, makes it pertinent to identify a means to distinguish salient features of any one particular profile from those of another.

All our previous publications on SDA applied to CCSN models have taken $V(r)$ to be an analytic, monotonic, smoothly-varying function of radius, { to start testing the inference procedure for parameter estimation.  This has helped us to adopt the technique to the neutrino collective oscillation problem.  Using an oversimplified matter potential helps us understand how applicable inference is to the collective oscillations problem.  It does not of course, adequately describe the turbulent supernova environment.}  In this work, we replace that form with one derived from an electron number density generated from a spherically symmetric (1D) hydrodynamical simulation of a CCSN environment~\cite{Radice:2017ykv, Wang:2024tbv}. This extension is noteworthy for two inter-related reasons.  First, the newly adopted matter profile is significantly more realistic: it is not necessarily monotonic, and may include discontinuities.  Second, it emerges from a significantly more physically detailed model.  Thus it brings our solutions closer to relevance for bearing upon a potential future CCSN detection.  Future work along this direction could involve extending this analysis to incorporate matter profiles generated by multi-dimensional simulations (e.g.,~\cite{nagakura2021core, vartanyan2019successful,Lentz:2015ApJ,Bollig:2020phc}), and possibly exploring the effects of trajectory/directional dependence.  { To be clear: we certainly do not claim that our extremely simpified CCSN model is a realistic representation of the phenomenon.  Rather, we are encouraged by our increemntal step toward realism, by virtue of adopting this new form for the matter potential.}

Within the context of simulated data from a CCSN model, our specific question in this paper is: What signature of the matter potential $V(r)$ within the CCSN envelope is contained in measurements of neutrino flavor made at physically accessible locations (i.e. in the vacuum-dominated regime)?  Specifically, can in-vacuum neutrino flavor measurements discriminate among different realistic matter potentials?  Indeed, we identify a reliable set of simulated measurements that can make this distinction.  We describe a straightforward metric that can be used to identify correct solutions, and discuss implications regarding a real detection.

\section{Model} \label{sec:model}

\subsection{\textbf{Neutrino flavor evolution}} \label{sec:modelForm}
Here we briefly describe the neutrino flavor model equations of motion, and refer the reader to our early publications (Refs.~\cite{armstrong2017optimization,armstrong2020inference,rrapaj2021inference}) for complete derivations.

The model is a single-angle, two-flavor treatment wherein two mono-energetic neutrino beams with different energies interact with each other and with a background of weakly charged nuclei, free nucleons, and electrons.  Immediately following core collapse, the neutrinos travel outward through the SN envelope, interacting with each other and with the dense matter potential surrounding the star immediately after core collapse.  The densities of the neutrinos dilute as some function of a position coordinate $r$, which we interpret as the distance from the neutrino-sphere in a supernova.  Importantly, the model contains coherent forward scattering only, and thus is solvable via traditional forward integration --- a consistency check for SDA solutions.

After decomposing the density matrices and Hamiltonians, respectively, into bases of Pauli spin matrices\footnote{The polarization vectors (Bloch vectors) are written in terms of neutrino density matrices: $\rho_i = \frac12(\mathbb{1}+\vec\sigma \cdot \vec P_i$). In the same manner, the Hamiltonian can be decomposed as $H_i = \frac12(\Tr(H_i) + \vec\sigma \cdot \vec V_i)$, where the $\vec V_i$ term contains contributions from vacuum oscillations, neutrino-matter interactions, and neutrino-neutrino interactions, as described in Eq.~\eqref{eq:model}.}, we write the flavor evolution of each neutrino in terms of a \lq\lq polarization vector\rq\rq\ $\vec P_i$ (see Ref.~\cite{Raffelt1993,Sigl1993} for details):
\begin{equation} \label{eq:model}
  \diff{\vec{P}_{i}}{r} = \left(\Delta_i \vec{B} + V(r) \hat{z} 
  +\mu(r) \sum_{j\neq i} \vec{P}_j \right) \times \vec{P}_i.
\end{equation}

In Equation~\eqref{eq:model}, $\Delta_{i} = \delta m^2/(2E_{i})$ are the vacuum oscillation frequencies of the two beams with energies $E_1$ and $E_2$, respectively, where $\delta m^2$ is the in-vacuum mass-squared difference.  The term $\vec{B}=\sin(2 \theta) \hat{x} -\cos(2 \theta) \hat{z}$ is a unit vector representing in-vacuum flavor mixing, where $\theta$ is the mixing angle between energy (mass) and flavor eigenstates.  Functions $V(r)$ and $\mu(r)$ are potentials for neutrino-matter and neutrino-neutrino coupling, respectively.  

The neutrino-neutrino coupling potential $\mu(r)$ is written as: 
\begin{equation} \label{eq:Vnu}
  \mu(r) = \frac{L_{\nu}}{4\pi \langle E_{\nu} \rangle R_{\nu}^2} \sqrt{2} G_F \left(1 - \sqrt{ 1 - \frac{R_\nu}{r}^2}\right)^2.
\end{equation}

\noindent Eq.~\eqref{eq:Vnu} reflects the manner in which coupling strength varies in the neutrino bulb model calculations that employ the single-angle approximation~\cite{duan2006simulation,duan2010collective}.  In our previous works, we have used a simplified functional form $\mu(r) \sim 1/r^4$, to which the above form reduces for $r \gg R_\nu$. The radial domains over which we examine neutrino flavor evolution in this study are sufficiently far from the proto-neutron star that we could have continued using this approximation.  We nevertheless chose to use the more general form here, with an eye on future works where we could potentially investigate flavor evolution closer to the PNS, in the \lq\lq fast\rq\rq\ oscillations regime.
{The model parameters taken to be constant and known are shown in Table~\ref{table:Known}.  

\setlength{\tabcolsep}{5pt}
\begin{table}[H]
\small
\centering
\begin{tabular}{|l |c | l | c |} \toprule
\hline
\textit{Parameter} & \textit{Value} &  \textit{Parameter} & \textit{Value} \\\midrule \hline
 $\Delta_1$ & 0.442 km$^{-1}$ & $\Delta_2$ & 0.562 km$^{-1}$ \\
 $\theta$ & 0.1 & $L_{\nu}$ & $1.23 \times 10^{52}$ erg/s \\
 $R_{\nu}$ & $20$ km & $\langle E_{\nu} \rangle$ & 13.5 MeV \\
\bottomrule \hline
\end{tabular}
\caption{\textbf{Model parameters taken to be known and fixed during the SDA procedure described in this paper.}  $\Delta_i$ are the neutrino vacuum oscillation frequencies, $\theta$ is the mixing angle in vacuum. } 
\label{table:Known}
\end{table}

\subsection{\textbf{The matter potential}}
\label{sec:modelMatterPotential}

The aim of this paper is to examine the SDA procedure's ability to handle more realistic matter profiles, compared to our previous work.  Importantly, we are interested in the procedure's ability to distinguish matter profiles that are relatively smoothly decaying from those containing shocks or other discontinuities - and to do so based on information contained in neutrino flavor measurements. 

To that end, we looked to a hydrodynamics simulation of a CCSN that explodes in one dimension.  The simulation is a 9.6-solar-mass model with a metallicity of $10^{-4}$~\cite{Radice:2017ykv, Wang:2024tbv}.  The output generated includes electron-fraction and electron number density profiles at eight distinct times post core bounce, as well as luminosities measured at a large distance from the source.  



For our purposes, the output of interest are the $n_e(r)$ profiles.  These values, measured in $cm^{-3}$, begin at the surface of the collapsing core and extend to 10,000 km.  The eight profiles were spaced at 100 ms: spanning 0.01 through 0.71 s post core bounce.  We chose four of the eight: those taken at 0.01, 0.31, 0.51, and 0.71 s.  We converted them to matter potentials $V(r)$ as seen by the neutrino field, by means of the relation:
\begin{equation*} \label{eq:Vm}
V(r)=\sqrt{2}G_F n_e\left(r\right).
\end{equation*}
\noindent This term embodies the neutrino mass Wolfenstein correction~\cite{wolfenstein1978neutrino}, as it arises from the coherent forward-scattering of neutrinos on the background electrons. 

For each of these matter profiles, we choose the starting point of our simulation grid to coincide with the matter density being $10^6$ g/cm$^3$. For spherically symmetric neutrino emission, flavor oscillations are not expected to be significant at larger densities due to matter-induced suppression---hence we adopt this density cutoff to help ease the computational burden on our simulations. Since the steepness of the matter profile changes with time, the radii at which the density falls to this threshold can be quite different across the various profile snapshots. This implies that the starting ratio of $V(r)$ to $\mu(r)$ can be quite different across the snapshots as well (since we use the same neutrino emission parameters throughout). In addition, the shape of the matter potentials across the four chosen snapshots are different -- the 0.01\,s snapshot does not exhibit a shock, whereas in the 0.31\,s, 0.51\,s, and the 0.71\,s snapshots a shock has formed that can be seen across the snapshots to be propagating forward on the one-dimensional grid (a sign of a successful explosion). Fig.~\ref{fig1} shows this set of four matter potentials, juxtaposed with the relative strengths of the other two terms of Eq.~\eqref{eq:model}: the vacuum oscillation frequencies ($\Delta_1$ and $\Delta_2$), and the neutrino-neutrino coupling term ($\mu$(r)).  
Hereafter the notation is as follows: the matter potentials at 0.01, 0.31, 0.51, and 0.71 s post core bounce are denoted: $V_{0.01}$, $V_{0.31}$, $V_{0.51}$, and $V_{0.71}$, respectively.

The nature of the ensuing neutrino oscillations could in principle depend on the hierarchy between the potentials, in addition to the shape of $V(r)$, and it is conceivable that measurements of neutrino flavor in the vacuum regime could contain imprints of these characteristics. In particular, as the matter potential decreases with radius, for each profile, the neutrinos experience an in-medium effective mass-level crossing, i.e., ``MSW resonance''~\cite{wolfenstein1978neutrino, mikheev1985resonance}, when $\Delta_i \cos{2\theta} \approx V(r)$ for neutrino $i$. If the variation of $V(r)$ is smooth and gradual across the MSW resonance region, then the neutrino traverses the resonance adiabatically -- meaning it remains in the same in-medium mass eigenstate as before and consequently experiences a near-complete flavor change. On the other hand, if the MSW resonance coincides with a shock, wherein $V(r)$ changes abruptly, then the evolution in non-adiabatic and the flavor conversion is less complete.  Thus, this set of four potentials will permit us to begin probing the SDA procedure's facility in distinguishing between matter profiles with different shock positions and different initial ratios of $V(r)$ to $\mu(r)$.  

\begin{figure*}[htb]    
\includegraphics[width=1.\textwidth]{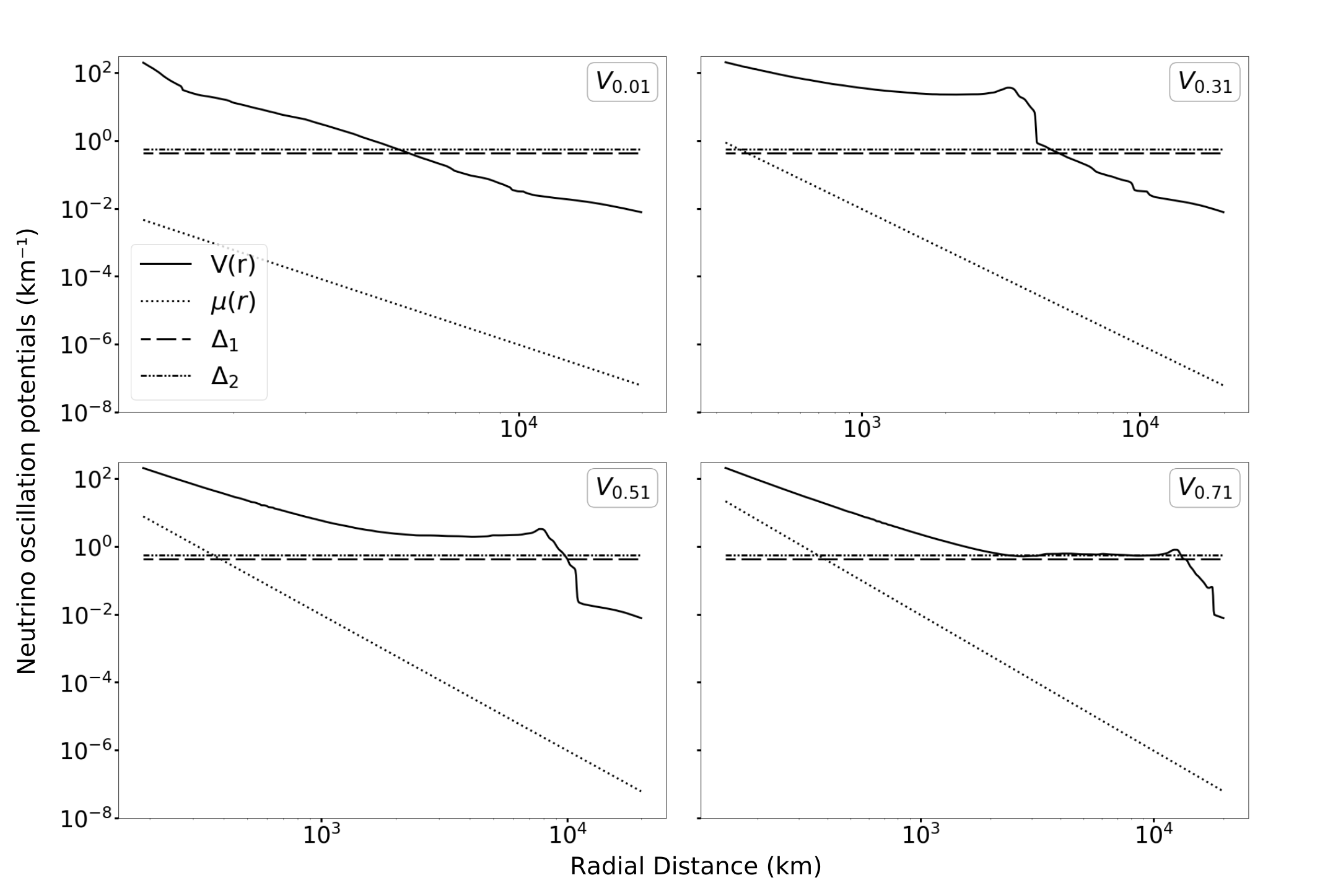}
\caption{\textbf{The matter potential $V(r)$, neutrino-neutrino coupling potential $\mu(r)$, and vacuum terms $\Delta_i$ of Eq.~\eqref{eq:model}, for each of the four matter potentials used in the SDA procedure.} 
The four matter potentials $V(r)$ shown here correspond to different snapshots of the simulation---0.01, 0.31, 0.51, and 0.71 s post core-bounce---labeled $V_{0.01}$, $V_{0.31}$, $V_{0.51}$, and $V_{0.71}$, respectively.  The slight differences across the four respective $x$-axes are a consequence of our use of a uniform cutoff in maximum matter density, $\rho = 10^6\,\text{g/cm}^3$, to determine the starting location of our domain in each case; see Sec.~\ref{sec:experDetails}.}
\label{fig1}
\end{figure*}

\subsection{\textbf{The measured model quantity}}

The simulated measurements in this paper are the $\hat{z}$ components of the neutrino polarization vector, $P_z$, for each neutrino beam. $P_z$ denotes the net flavor content of electron ($e$) flavor minus "$x$" flavor, the latter being a superposition of muon and tau flavors.  (Importantly: in considering an SDA procedure involving real Earth-based measurements, $P_z$ can be translated into a measured survival probability $P_\text{survival}$ according to: $P_{z} = 2 P_{\text{survival}}-1$ (see e.g.  Ref.~\cite{laber2023inference}.)


\section{Inference methodology} \label{sec:method}

\subsection{General formulation} \label{sec:methodGen}

Statistical data assimilation is an inference procedure wherein measured quantities are assumed to arise from the dynamics of a system that can be represented by a physical model.  SDA was designed for the case of sparse data, wherein only a subset of the model state variables can be associated with measurable quantities.  We write this model $\bm{F}$ in $D$ ordinary differential equations:
\begin{equation} \label{eq:ODE}
  \diff{x_a(r)}{r} = F_a(\bm{x}(r),\bm{p}(r)); \hspace{1em} a =1,2,\ldots,D,
\end{equation}
\noindent
where $r$ is the parametrization -- for example, distance or time.  The components $x_a$ of vector $\bm{x}$ are the model state variables, and $\bm{p}$ are unknown parameters to be estimated.  The parameters may be constants or may vary with parametrization $r$.  Based on measuring only $L$ of the $D$ total state variables (where $L$ is much less than $D$), we seek to predict the evolution of all state variables $D$, both measured and unmeasured.

\subsection{Optimizing a path integral} \label{sec:methodOpt}

We adopt a path-integral formulation\footnote{The path integral is an integral representation of the master equation for the stochastic process represented by Eq.~(\ref{eq:ODE})} of SDA, and seek the probability of obtaining a path $\bm{X}$ in the model's state space given observations $\bm{Y}$: 
\begin{align}
  P(\bm{X}|\bm{Y}) = e^{-A_0(\bm{X},\bm{Y})},
\end{align}
\noindent or: \textit{the path $\bm{X}$ for which the probability - given $\bm{Y}$ - is greatest is the path that minimizes the quantity $A_0$}, which we call our action.  Then, if we are able to find a computational form for $A_0$, we will be able to obtain the expectation value\footnote{Expectation values are the quantities of interest when the problem is statistical in nature.} of any function $G(\bm{X})$ on a path $\bm{X}$: 
\begin{align}
  G(\bm{X}) = \langle G(\bm{X}) \rangle = \frac{\int d\bm{X} G(\bm{X}) e^{-A_0(\bm{X},\bm{Y})}}{\int d\bm{X} e^{-A_0(\bm{X},\bm{Y})}}.  
\end{align}
\noindent For our purposes, the quantity of interest is the path itself: $G(\bm{X}) = \bm{X}$.  

The action can be written in two terms:
\begin{equation}
\begin{aligned}
\label{eq:action}
A_0(\bm{X},\bm{Y}) &= -\mathlarger{\sum} \log[P(\bm{x}(n+1)|\bm{x}(n))]\\ 
    &-\mathlarger{\sum} \text{CMI}(\bm{x}(n),\bm{y}(n)|\bm{Y}(n-1)).
\end{aligned}
\end{equation}

The first term describes Markov-chain transition probabilities governing the model dynamics.  The second term is the conditional mutual information (CMI) ~\cite{fano1961transmission}, which asks: "How
much information, in bits, is learned about event $\bm{x}$(n) upon observing event $\bm{y}$(n), conditioned on having previously observed event(s) $\bm{Y}$(n - 1)?”\footnote{For the reader interested in control theory: the measurement term can be regarded as a synchronization term, which are often introduced artificially into control problems.  Here, on the other hand, this term arises organically, via considering the impact of the information contained in those measurements.}.  See Ref.~\cite{abarbanel2013predicting} for a derivation of Eq.~\eqref{eq:action}.

The SDA problem is then cast as an optimization.  Here, the action -- the physical quantity we seek to minimize -- is a cost function.  Or: the cost function of the optimizer is equivalent to the action on paths in the state space that is searched.  Generally, the action surface is $((D + p) \times (N+1))$-dimensional, where $N+1$ is the number of discretized model locations, taken to be independent dimensions.  One seeks the path $\bm{X}^0 = \{ \bm{x}(0),\ldots,\bm{x}(N),\bm{p}(0),\ldots,\bm{p}(N) \}$ in state space on which $A_0$ attains a minimum value.  We extremize the cost function via the variational method: minima are found by requiring that small variations to the action vanish under small perturbations~\cite{oden2012variational}, thereby enforcing the Euler-Lagrange equations of motion upon any path.  
Many simplifications are made to write a computationally implementable form for $A_0$ (see Appendix A of Ref~\cite{armstrong2017optimization}).  Ultimately, the first and second terms of Eq.~(\ref{eq:action}) reduce to a \lq\lq model error\rq\rq $A_\text{model}$ and \lq\lq measurement error\rq\rq\ $A_\text{meas}$, respectively~\cite{abarbanel2013predicting}:
\begin{widetext}
\begin{equation} \label{eq:actionlong}
\begin{split}
A_0 =& R_f A_\text{model} + R_m A_\text{meas}\\
A_\text{model}=&\frac{1}{{N}D}	\mathlarger{\sum}_{n \in \{\text{odd}\}}^{N-2} \, \mathlarger{\sum}_{a=1}^D \\ 
   & \Bigg[ \left\{x_a(r_{n+2}) - x_a(r_n) - \frac{\delta r}{6} [F_a(\bm{x}(r_n), \bm{p}(r_n)) + 4F_a(\bm{x}(r_{n+1}),\bm{p}(r_{n+1})) + F_a(\bm{x}(r_{n+2}),\bm{p}(r_{n+2}))]\right\}^2 \\
   & + \left\{ x_a(r_{n+1}) - \frac12 \left(x_a(r_n)+x_a(r_{n+2})\right) - \frac{\delta r}{8} [F_a(\bm{x}(r_n),\bm{p}(r_n)) - F_a(\bm{x}(r_{n+2}),\bm{p}(r_{n+2)})]\right\}^2 \Bigg] \\
  A_{\text{meas}} =& \frac{1}{N_{\text{meas}}} \mathlarger{\sum}_{r_m \in \{\text{meas}\}} \, \mathlarger{\sum}_{l=1}^d  \left[\left(y_{l}\left(r_m\right) - h_{l,m}(\bm{x}(r_m) \right)^2 \right]
\end{split}
\end{equation}
\end{widetext}

Model error $A_\text{model}$ quantifies the divergence of the prediction from model dynamics, for all $D$ state variables $x_a$ -- both measured and unmeasured.  The outer sum on $n$ runs through all odd-numbered discretized locations.  The sum on $a$ runs through all $D$ state variables. The terms within the first and second sets of curly brackets represent the errors in the first and second derivatives of the state variables, respectively.

Measurement error $A_\text{meas}$ quantifies the divergence of the prediction from any measurements $y_l$ that had been furnished to the procedure. The variables $y_l$, for $l=1,\ldots,d$, represent the $d$ quantities measured at locations $r_m \in \{{\text{meas}}\}$, where $N_\text{meas}$ is the total number of locations. These values are to be compared against the quantities $h_{l,m}(\bm{x})$, where $h_{l,m}$ are transfer functions that relate the quantities being measured to the associated model state variables, at each location. For our simulations, the measured quantity is $P_z$ for each neutrino energy; thus the transfer function $h$ is simply 1.0.   

\subsection{\textbf{Avoiding multiple minima}} \label{sec:methodAnnealing}

A nonlinear model will have an action surface that is non-convex; i.e. it will possess multiple minima.  To identify a "lowest" minimum, we iterate in terms of the ratio of model and measurement error, $R_f$ and $R_m$, respectively\footnote{Generally, $R_m$ and $R_f$ are inverse covariance matrices.  We take the measurements to be mutually independent, rendering the matrices diagonal.}~\cite{ye2015systematic}.  This "annealing" procedure goes as follows.

First, $R_m$ is taken to be a constant (in this paper it is set to 1.0), and we write $R_f$ as: $R_f = R_{f,0}\alpha^{\beta}$, where $R_{f,0} = 10^{1}$, $\alpha = 2.0$, and we initialize our "annealing parameter" $\beta$ at zero.  Relatively free from model constraints, the action surface is essentially convex, and we can readily calculate a single solution to the problem, which corresponds to our first estimate of the action $A_0$.  Of course, in this landscape, the model dynamics have yet to be captured -- but this is a critical starting point. 

Next, beginning our second search at the location of the first estimate for $A_0$, we increase $\beta$ slightly, to impose weak model constraints -- and we obtain an updated estimate of $A_0$.  We do this recursively, each time recalculating $A_0$, toward the deterministic limit of $R_f \gg R_m$.  The aim is to remain sufficiently near to the lowest minimum as the dynamics grow increasingly well resolved, so as not to become trapped in a local minimum along the way\footnote{The complete procedure -- a variational approach to extremization and an annealing method to identify a lowest minimum of the cost -- is termed variational annealing (VA).}.

We have shown~\cite{armstrong2020inference} that a plot of the Action as a function of annealing parameter $\beta$ offers a litmus test for the correct solution.  Namely: the best predictions of neutrino evolution correspond to paths of lowest action.  This will be a critical tool for interpreting the results presented in this paper.

\section{Task for the SDA procedure} \label{sec:expers}

Given the dynamical model of Eq.~\eqref{eq:model}, we ran four independent sets of optimization designs, each taking as the matter potential of Eq.~\eqref{eq:model} one of the set of four depicted in Fig.~\ref{fig1}.  

Then, within each of those four sets, we ran four independent optimization designs, distinguished by the specific simulated neutrino flavor measurements to be optimized with the model.  In the first design, the neutrino measurements to be optimized with the model were those measurements that indeed had been generated by that particular model.  In the other three designs, the neutrino measurements taken were those generated by a model containing one of the other three potentials --- that is: deliberate mismatch.  Our aim was to ascertain whether the SDA procedure could identify the former ("correct") pairing versus the latter three ("incorrect") pairings.

As an example for clarification: in the first set of optimizations, we took the matter potential of Eq.~\eqref{eq:model} to be $V_{0.01}$.  We first optimized that model with the in-vacuum neutrino flavor measurements generated by a model that indeed contained $V_{0.01}$ (the "correct" pairing).  Then we optimized the model containing $V_{0.01}$ with measurements generated by a model containing: i) $V_{0.31}$; ii) $V_{0.51}$; and iii) $V_{0.71}$.  So, the four optimizations for each potential included one "correct" and three "incorrect" pairings.  Across the four potentials, then, we ran sixteen designs total.

As simulated measurements, we furnished to the procedure the final 1,000 values of $P_z$ for each neutrino beam, out of the total $\sim$ 100,000 radial locations on the discretized path.  These measurement locations are in the vacuum-dominated regime, intended to simulate Earth-based observation\footnote{The physical justification behind providing 1000 in-vacuum measurements as opposed to just one is that, in reality, neutrinos experience kinematic decoherence by the time they arrive at the earth. As a result, their flavor is no longer evolving at that point, making it easier to extract information about their flavor composition (because the oscillation phases have damped out). Providing multiple in-vacuum measurements to the SDA procedure is a way to effectively mimic this information, without explicitly adding decoherence to the model itself.}.  Our question for the SDA procedure, then, was the following: \textit{do these sparse in-vacuum measurements of $P_z$ contain sufficient information for the SDA procedure to distinguish between a "correct" and "incorrect" pairing of model versus measurement?}  That is: we challenged the SDA procedure to take these sparse in-vacuum flavor measurements and \textit{infer} the shape of the matter potential that the neutrinos had encountered at much earlier radii within the matter-dominated regime of the CCSN envelope.

For the measurements to be furnished to the SDA procedure, we chose the value of $P_z$ for each beam at 1,000 discretized locations in the vacuum-dominated regime. (We remind the reader that $P_z$ can be translated to a measurable Earth-based survival probability.)  To generate these simulated measurements, we used a forward-integration code, initializing both beams as pure electron-flavor at emission (i.e. the $P_z$ component was set to +1.0 for each), evolving the full model (i.e. $P_x$, $P_y$, and $P_z$ for both beams) through the MSW resonance, and then through an appreciable distance where the in-vacuum potential dominates (recall the relative scalings depicted in Fig.~\ref{fig1}.  This discretized path contained $\sim$ 100,000 locations (the precise number varied across the potentials; see Sec.~\ref{sec:experDetails}).  We permitted the SDA procedure access to only the final 1,000 values of $P_z$ for each beam - locations that resided well within the vacuum-dominated region.  Fig.~\ref{fig3} shows a schematic of the design for optimization.  The Caption is intended to provide a distillation of the in-text explanation.

\begin{figure*}[htb]
    \includegraphics[width=0.8\textwidth]{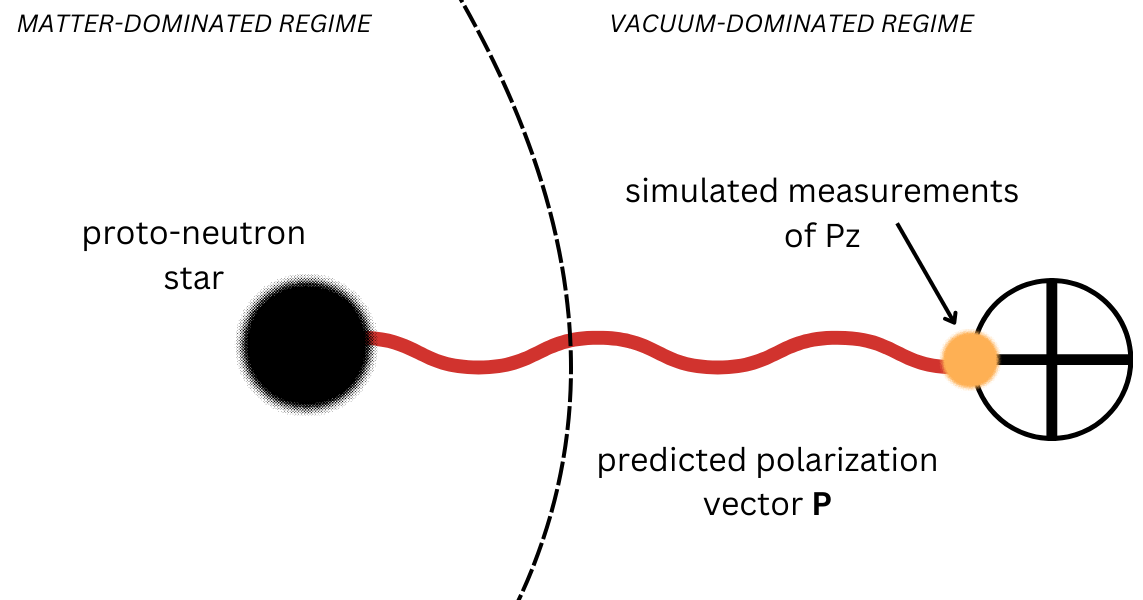}
    \caption{\textbf{Schematic of the optimization design.}  Simulated measurements of the $P_z$ components of the neutrino flavor polarization vectors $\bm{P}$, for both beams, are provided to the SDA procedure in a small range in-vacuum only (orange circle).  The procedure is then challenged to predict the evolution of the full polarization vectors, both within that region in which measurements are provided, and backward to the source (red line).}
    \label{fig3}
\end{figure*}

\subsection{\textbf{Details of the procedure}} \label{sec:experDetails}
The simulated measurements were generated by Python, using an integrator (odeint) with an adaptive step.  For comparison with the results from the optimization method, the output of this code was saved on a uniform grid with a spacing of $0.2\,\text{km}$ between successive grid-points.

We performed the optimization with the open-source Interior-point Optimizer (Ipopt)~\cite{wachter2009short}.  Ipopt employs a Hermite-Simpson method of discretization and a constant (non-adaptive) step.  The reader may note the slightly differing $x$-axes across the four panels of Fig.~\ref{fig1}.  The number of discretized steps differed slightly for each, because we chose to cut each matter potential off above a certain maximum density, i.e., distance from the source.  Our motivation was computational limitation: the SDA procedure has difficulty handling rapidly-changing derivatives, due to its non-adaptive step.  Moreover, as stated earlier, one does not expect significant neutrino oscillations to occur anyway above those matter densities, in spherically symmetric models.  For $V_{0.01}$, $V_{0.31}$, $V_{0.51}$, and $V_{0.71}$, we began our record of the neutrinos' outward trajectory at $\simeq$ {1204, 324, 189, and 146 km,} respectively, from the source.  Across all optimization designs, we used a uniform step size $\delta r$ of 0.2 km.

For each of the sixteen optimization designs described above, we initialized ten independent trials.  Each trial was distinguished by a randomly-generated set of initialized guesses for values of the state variables at each location on the discretized one-dimensional path in the state-and-parameter space.

The discretization of state space, calculations of the model Jacobean and Hessian matrices, and the annealing procedure were done via a Python interface~\cite{minAone} that generates C code that Ipopt reads.  Simulations were run on a computing cluster equipped with 201 GB of RAM and 24 GenuineIntel CPUs (64 bits), with 12 cores each.  All our procedures are available in the open-access Github repository of Ref.~\cite{github}.

\section{Result} \label{sec:result}

We remind the reader that ultimately we will examine optimization problems that are not readily solvable via forward integration, and thus we will not have a point of comparison for SDA predictions.  In such a case, we will seek to trust the SDA procedure in leading us to the physical solution.  Thus we require a litmus test for an optimal solution.  In Ref.~\cite{armstrong2020inference}, we verified that such solutions -- those that are most compatible with model and measurements -- were those corresponding to paths of least action\footnote{In addition, that reference showed that the action level can also be used to rule out parameter regimes in which correct solutions are not possible.}.  We shall now apply that finding to analyze the optimization results.

\subsection{\textbf{Key findings}}

Key results are as follows:
\begin{itemize}
    \item Using the 1,000 in-vacuum $P_z$ measurements, the SDA procedure is able to discriminate between "correct" and "incorrect" pairings of matter potentials versus measurements\footnote{A preliminary study of shifting that range of 1,000 locations to different specific regions within the vacuum region indicates that the results are unchanged -- as in principle they should be.}.  Here, the embodiment of this discrimination is in the action as a function of annealing parameter $\beta$.
    \item Specifically: the indication of an "incorrect" pairing is that the action increases exponentially at high values of $\beta$ -- that is, as it enters the model-dominated (as opposed to the measurements-dominated) regime.  In this regime, the SDA procedure has been furnished with sufficient model resolution to recognize the incompatibility of model with the measurements -- \textit{even though the measurements were only provided in the vacuum region}.
    \item The results for all 16 pairings (four "correct" and 12 "incorrect") are consistent across ten trials generated for each.  Here, each trial was defined by a distinct randomly-generated initialized search location in the state-and-parameter space.  This we take as a measure of the results' robustness.
\end{itemize}

\subsection{\textbf{The action as a function of annealing}}
\begin{figure*}
    \includegraphics[width=1.\textwidth]{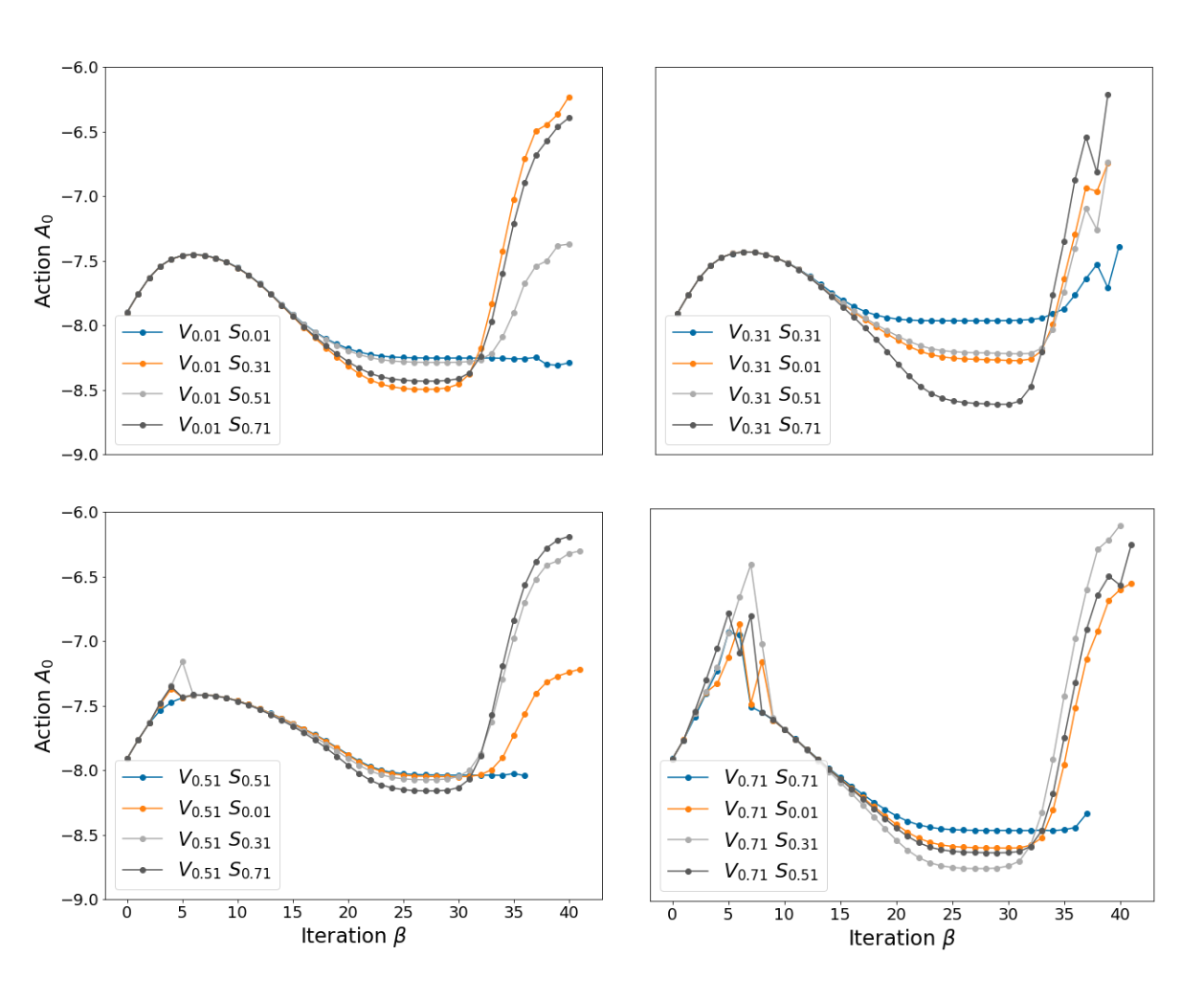}
    \caption{\textbf{Action as a function of annealing parameter $\beta$}, for all 16 SDA optimization designs described in Sec.~\ref{sec:expers}, and depicted in Fig.~\ref{fig3}.  The action is the cost function, which consists of both measurement and model error; $\beta$ tracks the gradually-increasing relative weight of the model error.  The solution corresponding to the lowest action in the model-dominated regime (i.e. high $\beta$) is expected to be optimal.  Note two features of the plots: a plateau in the measurement-dominated regime, and -- for the "incorrect" pairings -- a steep rise in the model-dominated regime.  The notation "$V_j S_k$" indicates: a model containing Matter Potential $V_j$ is optimized with simulated measurements generated using a model containing matter potential $V_k$, where $\{ j,k \} \in \{0.01, 0.31, 0.51, 0.71 \}$.}
    \label{fig4}
\end{figure*}

Fig.~\ref{fig4} shows the action, over the course of annealing, for all 16 optimization designs described in Sec.~\ref{sec:expers}\footnote{Each of these 16 action plots is representative of the ten randomly-initialized trials from which it was selected; see Sec.~\ref{app:all10ICs}.}.  To remind the reader: the annealing procedure takes us gradually from the regime in which the deviation from the furnished measurements ("measurement error") dominates the action to one in which the deviation from model dynamics ("model error") dominates (recall Sec.~\ref{sec:methodAnnealing}.

The panels of Fig.~\ref{fig4} correspond to the matter potentials at 0.01\,s (top left), 0.31\,s (top right), 0.51\,s (bottom left), and 0.71\,s (bottom right) post core bounce (the same ordering as Fig.~\ref{fig1}).  Within each panel, the legend notation "$V_jS_k$" indicates that Matter Potential $V_j$ has been optimized with simulated measurements $S_k$ that were generated using a model containing Matter Potential $V_k$, where $\{ j,k \} \in \{0.01, 0.31, 0.51, 0.71 \}$.  The blue path corresponds to the "correct" pairing, and the other three paths correspond to the "incorrect"; see the legends for color assignments.  

Note two features of all four panels in Fig.~\ref{fig4}: 1) a "plateau" of the action around $\beta$ 20-30, and 2) for the "incorrect" pairings: a rapid rise in action value beyond the plateau.

Feature 1, the action "plateau," shows the optimization still living in the measurements-dominated regime, and apparently settled on an optimal value of the action.  In this regime, the model is not yet sufficiently resolved for the SDA procedure to identify compatibility (or lack thereof) with the sparse in-vacuum measurements that were provided.  (The reader may note that some of the "incorrect" optimization designs yield action plateaus at lower values than some of the "correct" optimization designs.  This is not significant; see Appendix~\ref{app:Plateaus}.)

As annealing parameter $\beta$ continues to increase, however, the model-error weight of the action increasingly dominates.  Eventually, any discrepancy between the provided in-vacuum $P_z$ measurements and the model dynamics grows apparent.  This brings us to Feature 2 of the action plots: the rapid rise in action value from the plateau, for the "incorrect" pairings.  For each of the four matter potentials examined: this discrepancy becomes apparent vastly faster for the "incorrect" pairings than for the "correct" ones\footnote{In fact, for the $V_{0.01}$ and $V_{0.51}$ cases, the "correct" pairing remains stable on the plateau.  For the $V_{0.31}$ and $V_{0.71}$ "correct" pairings, there is a slight rise in action at high values of beta, which may be due to discretization error; see Sec.~\ref{app:Plateaus}.}.

\subsection{\textbf{Interpreting the action plots via the SDA solution}}

Our interpretation of Fig.~\ref{fig4} derives from examining the quality of the SDA's match to the state variables' evolution, compared to the evolution generated by the forward integration.  Fig.~\ref{fig5} shows this comparison, for one "correct" (left) and one "incorrect" (right) pairing.  The y-axes are $P_x$, $P_y$, and $P_z$, for Neutrino Beams 1 and 2.  The "correct" pairing at left is the optimization of a model containing $V_{0.71}$ with measurements generated by that same potential.  The "incorrect" pairing at right is the optimization of a model containing $V_{0.01}$ with measurements generated by a model containing $V_{0.71}$. The SDA solutions in both cases are shown in red, and they are each juxtaposed against the forward integration solution obtained from $V_{0.71}$ -- since in each case, the simulated measurements provided to the optimization model were from the $V_{0.71}$ simulation.
We chose this particular pairing for the striking visual contrast between the relatively smooth and monotonically decaying 0.01\,s profile, compared to the 0.71\,s profile where $V(r)$ remains close to resonance over an extended region in $r$.
The region in which the simulated measurements of $P_z$ were given to the procedure is denoted by an orange vertical bar.  

The "prediction" region -- that is, the region in which the procedure was challenged to \textit{infer} the model evolution history, occurs at all points to the left of that bar.  For the "correct" pairing, the optimization result shows excellent agreement with the result from forward integration - as expected.  For the "incorrect" pairing, there is significant discrepancy - also as expected.  See the caption of Fig.~\ref{fig5} for details.

Now, these results correspond to the model-dominated regime of the SDA procedure, i.e., at a value of the annealing parameter $\beta = 36$, 
where the action plots of Fig.~\ref{fig4} rise steeply for "incorrect" pairings, indicating that the SDA procedure has recognized a discrepancy between model and measurements.  This is not reflected by-eye in Fig.~\ref{fig5} within the measurement region.  So let us zoom in.
\begin{figure*}[htb]
    \includegraphics[width=1.\textwidth]{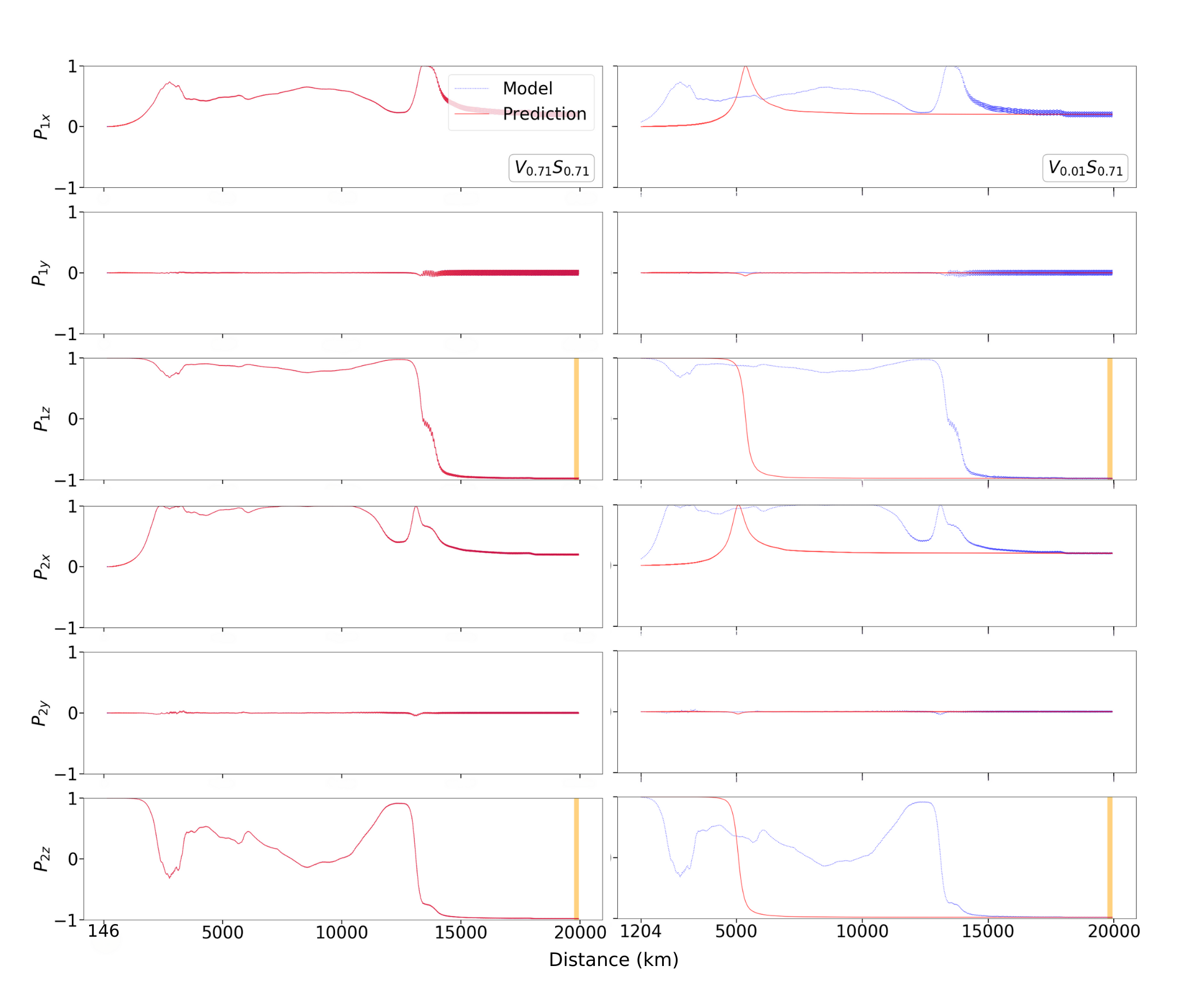} 
    \caption{\textbf{Forward-integration (blue) versus SDA (red) solution of model state variables}, for two optimizations of model and measurements.  Rows are the polarization vector components for beams 1 and 2, respectively; from top: $P_{1x}$, $P_{1y}$, $P_{1z}$, $P_{2x}$, $P_{2y}$, $P_{2z}$.  The vertical orange strip indicates the location where measurements were made of $P_z$.  In both sets of panels (\textit{left} and \textit{right}), the curves labeled ``Model'' (blue) are generated via forward integration using the matter potential $V_{0.71}$.  \textit{Left panels, red:} an optimization prediction obtained using a "correct" pairing of Matter Potential $V_{0.71}$ with measurements generated by that same potential (this corresponds to the action path "$V_{0.71}S_{0.71}$" in Fig.~\ref{fig4}).  The agreement with the model is perfect by-eye across the full domain.  \textit{Right panels, red}: an optimization prediction obtained using an "incorrect" pairing of Matter Potential $V_{0.01}$ with measurements generated by the $V_{0.71}$ matter potential (this corresponds to the action path "$V_{0.71}S_{0.01}$" in Fig.~\ref{fig4}).  As expected, disagreement between the prediction that uses $V_{0.01}$ and the model that uses $V_{0.71}$ is obvious by-eye across a majority of the domain. The important question then is whether this prediction using $V_{0.01}$ can nonetheless be reconciled with the model of $V_{0.71}$ \textit{at the locations where the measurements are made}; see Fig.~\ref{fig6}. Both predictions correspond to the model-dominated regime ($\beta=36$) in Fig.~\ref{fig4}.}
    \label{fig5}
\end{figure*}

\begin{figure*}[t]
    \includegraphics[width=1.\textwidth]{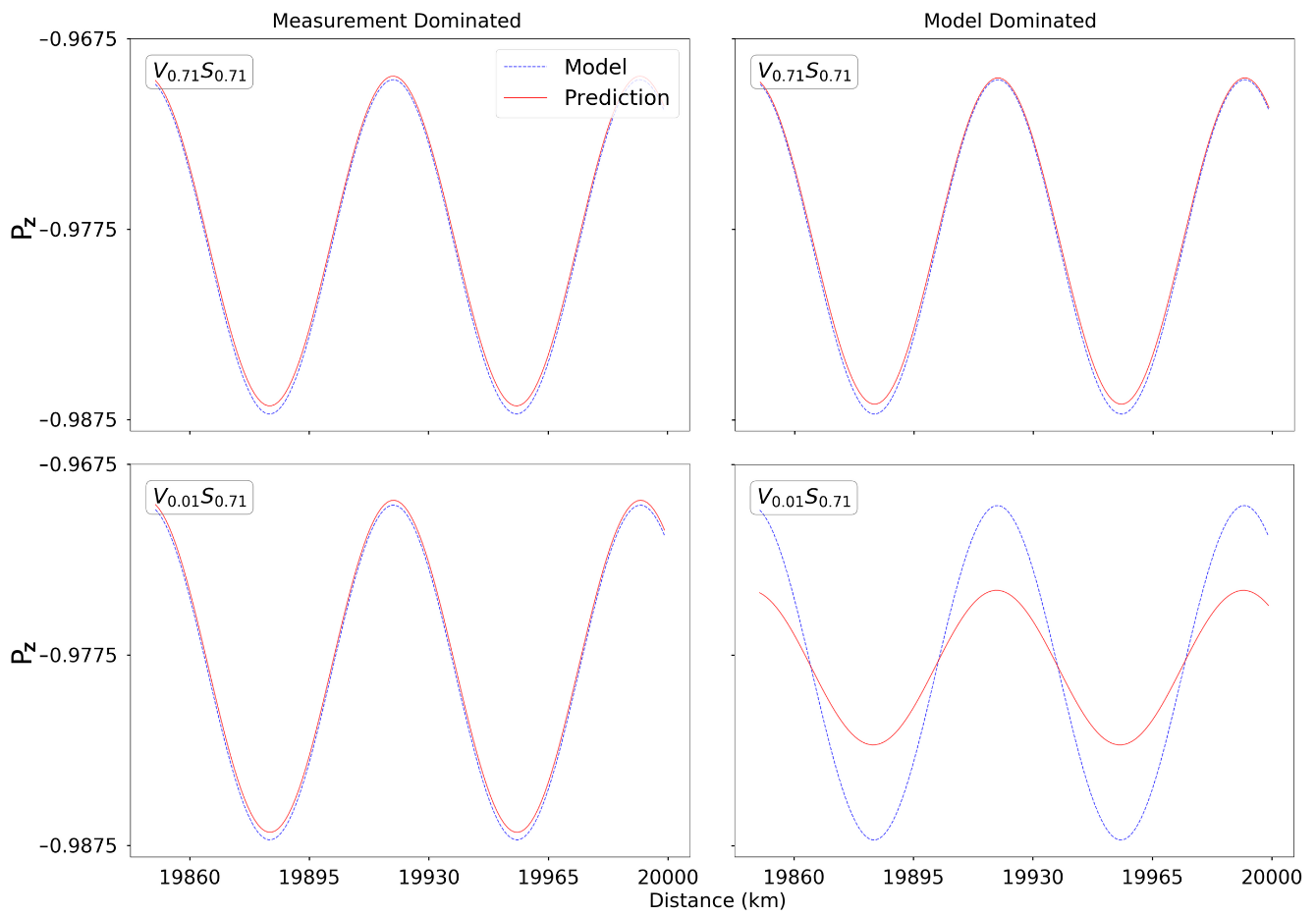}
    \caption{\textbf{Forward-integration versus SDA solution of model state variables} for Neutrino Beam 1, for the two optimization designs of Fig.~\ref{fig5}, now zoomed in on the $P_z$ measurements in the vacuum region.  \textit{Left} and \textit{right} panels compare the result corresponding to a measurements-dominated action (i.e. the "plateau" region of the action, at $\beta=27$) versus a model-dominated action (beyond the plateau, at $\beta=36$). In each panel, the curves labeled ``Model'' (blue) are generated via forward integration using the matter potential $V_{0.71}$. \textit{Top}: an optimization prediction obtained using a "correct" pairing of Matter Potential $V_{0.71}$ with measurements generated by that same potential.  Note that the agreement between prediction and model remains intact at high model resolution.  \textit{Bottom}: an optimization prediction obtained using an "incorrect" pairing of Matter Potential $V_{0.01}$ with measurements generated by Potential $V_{0.71}$.  The agreement between prediction and model worsens at high model resolution, as expected.  This behavior is reflected in the behavior of the action plots of Fig.~\ref{fig4}: the action path corresponding to the "correct" pairing ("$V_{0.71}S_{0.71}$") remains near the plateau value even at high $\beta$, while the action corresponding to the "incorrect" pairing ("$V_{0.01}S_{0.71}$") rises exponentially with increasing $\beta$.  (The result for Neutrino Beam 2 is comparable; not shown.)}
    \label{fig6}
\end{figure*}

Fig.~\ref{fig6} shows the same two optimization designs of Fig.~\ref{fig5}, now zoomed in on the final 150 locations at which measurements were provided -- a region corresponding to the vacuum regime.  The region is shown at two values of annealing parameter $\beta$.  At top, at $\beta=27$, we are in the plateau region of the action plots of Fig.~\ref{fig4}, before the procedure has realized anything amiss between model and measurements.  At bottom, we have reached $\beta=36$ -- the model-dominated regime wherein the "incorrect" pairings rise steeply in action in Fig.~\ref{fig4}.  
 
Note that in Fig.~\ref{fig6}, for the "correct" pairing (left), the match between forward-integration and SDA solution is excellent -- in both the measuremements- (top) and model-dominated (right) regimes, indicating high compatibility between model and measurements even at high model resolution.  For the "incorrect" pairing (right), however, the match between forward-integration and SDA solution is notably worse at high (bottom) compared to low (top) model resolution.  This is reflected in the action values of Fig.~\ref{fig4} corresponding to "incorrect" pairings, which rise steeply from the plateau.

Now, imagine the more realistic case wherein we do not know what the solution via forward integration looks like.  In that case, we will not have a point of comparison against the SDA prediction -- and we hope that the reader now grasps the utility of plotting the action as a function of annealing parameter $\beta$.  While we may not always be able to integrate forward from known initial conditions, we will be able to formulate a means to calculate the action.  And here we have endeavoured to show that that value correlates directly with the quality of prediction.  It is in this way that we anticipate the SDA procedure will ultimately lead us to new physics that the technique of forward integration is unable to access.

\section{Discussion} \label{sec:disc}

We have shown, within the confines of a small-scale model, that in-vacuum neutrino flavor measurements contain a signature of the matter potential profile that those neutrinos encountered at vastly earlier radii inside the CCSN envelope, and that we are able to use the behavior of the action value over an annealing procedure as a proxy to infer that signature.  These results have implications for interpreting features of a real galactic CCSN signal based on Earth-based neutrino survival probabilities.

{The four $V(r)$ snapshots have the following differences relative to one another: (i) different initial ratios of $V(r)$ to $\mu(r)$ to $\Delta_i$; and (ii) different locations of the shock (or its absence altogether, in the case of $V_{0.01}$) in relation to the location of the MSW resonance. In principle, each of these could be imprinted in the flavor evolution of the neutrinos. The simulated measurements available to the procedure were values of $P_z$ in the vacuum regime (at $r \sim 20000\,\text{km}$). Multiple $P_z$ measurements in this region are able to furnish information about the amplitude, frequency, and vertical offset of the $P_z$ waveforms, which in turn are determined by the flavor evolution histories prior. For the particular scenario examined in this work, wherein the initial conditions are $P_{z,i} = +1$ for both neutrinos, we were able to verify that the initial ratio of $V(r)/\mu(r)$ or $\mu(r)/\Delta_i$ does not significantly affect the final outcome, in terms of amplitude or vertical offset of $P_z$ in the vacuum regime (this was done by comparing the forward integration results for $\mu(r) = 0$ with those obtained using $\mu(r)$ from Eq.~\eqref{eq:Vnu}). This led us to conclude that the simulated measurements furnished to the SDA procedure in our chosen setup would only contain information about the interplay between the shock and the MSW resonance.}

{Among the four snapshots, the one without the shock (0.01\,s) results in flavor evolution that is the most adiabatic across the MSW resonance, resulting in the smallest final amplitude of $P_z$. In contrast, the other three snapshots have the shock coinciding (or nearly so) with the MSW resonance region, resulting in a partial loss of adiabaticity and a larger oscillation amplitude of $P_z$ in the vacuum regions (and the amplitudes are different across the three shocked profiles well). As evident from our results, the SDA procedure is able to use this information contained in the simulated measurements to differentiate among the different snapshots, i.e. distinguish a "correct" pairing of measurements and $V(r)$ apart from an "incorrect" pairing.
}


\section*{Appendix A: Details on the behavior of the Action over annealing} \label{app:Plateaus}

\noindent \textbf{Value of action across "correct" and "incorrect" pairings}\\

The reader might note in Fig.~\ref{fig4} that some of the action plateaus corresponding to "incorrect" pairings are lower in value than some "correct."  This is not concerning, and here we explain why.

Fig.~\ref{figAppA} shows that there exists intrinsic variation in the values of the action plateaus across the four "correct" pairings themselves.  This may be due the differences in sampling resolution, to be discussed below.

Note that this spread in action plateau values across the "correct" pairings is roughly the size of the spread across the action plateaus for each of the four sets of pairings shown in Fig.~\ref{fig4}.  Thus, the fact that some "incorrect" pairings yield lower action plateaus than "correct" pairings is not significant for our purposes.  Moreover, in the plateau region, the procedure simply does not yet have sufficient model resolution to distinguish "correct" from "incorrect."  \\
\begin{figure}
    \includegraphics[width=0.5\textwidth]{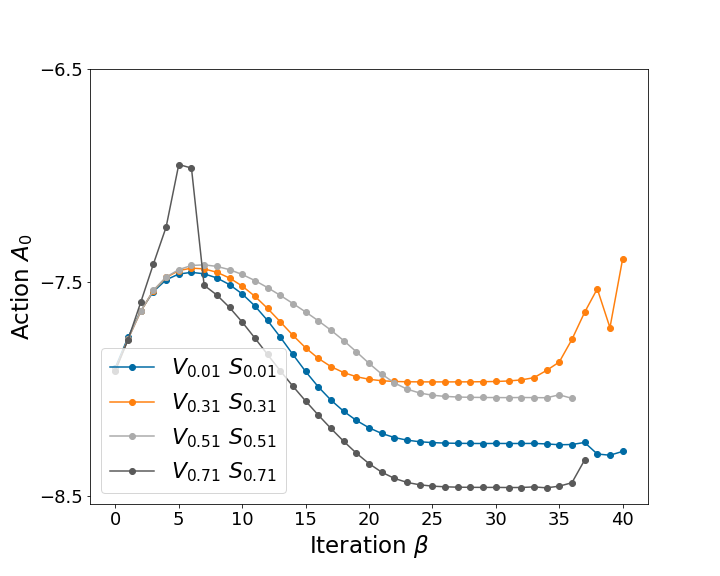}
    \caption{\textbf{Action as a function of annealing parameter $\beta$ for the four "correct" pairings of matter potential and measurements.}  That is: matter potential $V_{0.01}$ was optimized with flavor measurements produced by a model that indeed contained that $V_{0.01}$ potential, and so on, for the other three potentials.}
    \label{figAppA}
\end{figure}

\noindent \textbf{Behavior of action on the "plateaus" for "correct" pairings}\\

Fig.~\ref{fig4} shows that the "correct" pairings for the 0.31- and 0.71-s matter potentials show slight increases in action at high values of $\beta$ (compared to the 0.01- and 0.51-s matter potentials that remain on the plateau) -- albeit these rises are slower by orders of magnitude than the rises for "incorrect" pairings.  Indeed, one should not necessarily expect the "correct" pairings to remain strictly on the plateaus. 
 
As model weight increases, one expects to ultimately encounter a "discretization error."  This refers to an increasing discrepancy between the forward-integration versus the SDA optimization solution at high model resolution, even in the limiting case where perfectly accurate measurements are provided at all locations.  Discretization error is due to the fact that the two methodologies employ different means of discretizing the state-and-parameter space (as described in Sec.~\ref{sec:experDetails}).  As the model dynamics become increasingly well resolved, this discrepancy is inevitably revealed. 

Now, we might have expected the simulations generated by the 0.31- and 0.71-s matter potentials to be more sensitive to discretization error, compared to those generated by the 0.01- and 0.51-s matter potentials, for the following reason: the former contain regions of significantly faster flavor oscillations {prior to the MSW resonance regions, even though the amplitudes of these oscillations may not be substantial}.  And our choice of constant sampling across all potentials (Sec.~\ref{sec:experDetails}) was such that the fastest flavor oscillations that arose from the 0.31- and 0.71-s potentials were sampled roughly eight to ten times per oscillation, while the fastest oscillations arising from the 0.01- and 0.51-s potentials were sampled 20 to 30 times per oscillation.  This difference in resolution could account for the fact that the latter two remained strictly on the plateau while the former two rose slightly, even at the same model weight.

The interesting physical question here is: \text{why} the differences in oscillation frequency across the four potentials?  That question is beyond the scope of this paper, but may be related to the relative strengths of the matter potentials compared to the neutrino-neutrino and vacuum terms depicted in Fig.~\ref{fig1}, as well as the specific location of the shock as it travels outward in time; see Sec.~\ref{sec:disc}. It is interesting to note that the 0.31\,s and 0.71\,s profiles are also those with the highest $P_z$ oscillation amplitude in the vacuum regime. However, that is entirely a consequence of $V(r)$ (shock-MSW interplay) and not $\mu(r)$ [as noted in \textit{Discussion}], so it may not be related to their respective discretization errors in the matter-dominated regime.

\section*{Appendix B: consistency across trials, for each experiment} \label{app:all10ICs}

For each of the 16 optimization designs depicted in Fig.~\ref{fig4}, we generated ten trials.  Each trial was defined by a different randomly-generated set of initial guesses regarding the values of all state variables at each of the $\sim$ 100,000 locations on the discretized path in the state space.  The 16 action plots shown in Fig.~\ref{fig4} are those containing the lowest value of the action at a value of $\beta$ of 36 (i.e. in the model-dominated regime) within their respective sets of ten.  Across all 16 sets, the standard deviation in the logarithm of the action is one part in $10^4$.  Fig.~\ref{figs_appB} offers a depiction.  Thus, we considered the 16 paths displayed in Fig.~\ref{fig4} as representative of the full study. 

\begin{figure*}[h]
    \includegraphics[width=1\textwidth]{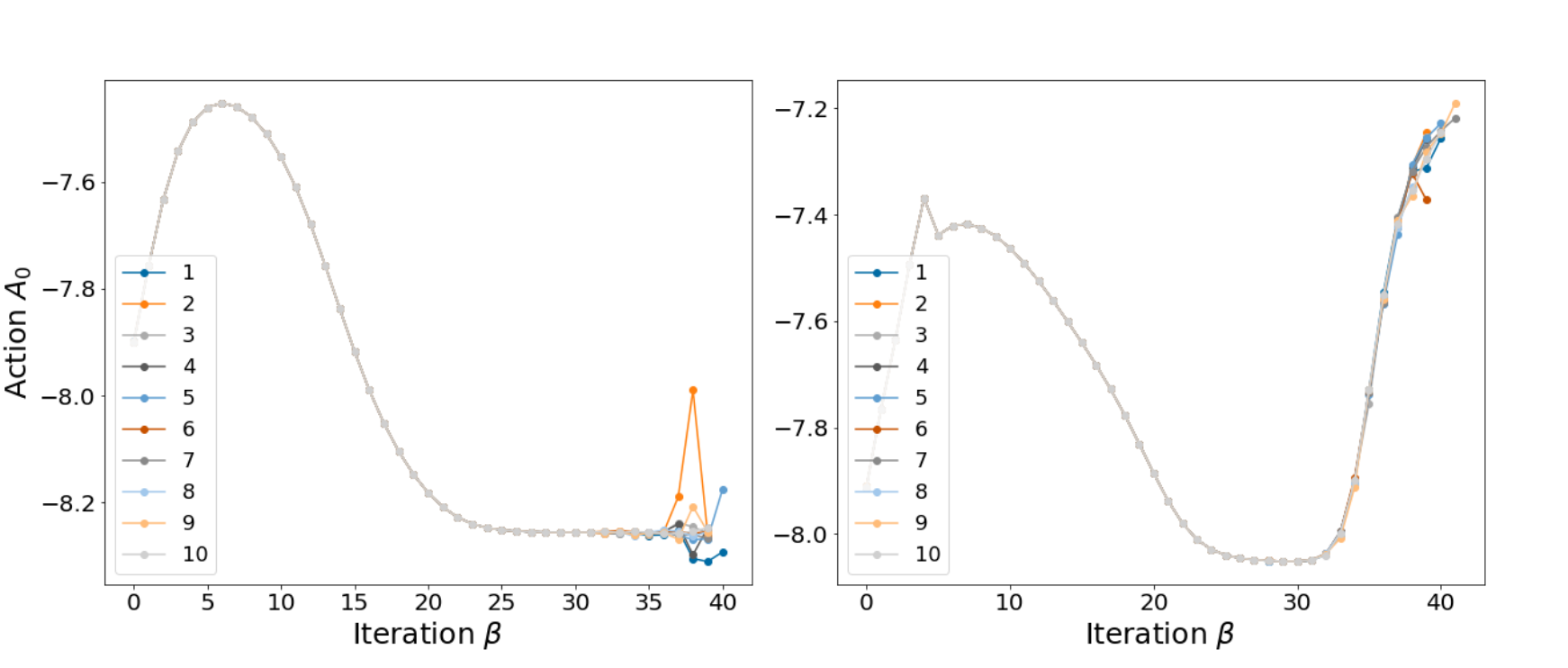}
    \caption{\textbf{Action as a function of annealing parameter $\beta$ across all ten randomly-initialized trials, for one "correct" pairing (left: $V_{0.01}S_{0.01}$) and one "incorrect" (right: $V_{0.51}S_{0.01}$)}.  Within each set of ten, we chose the path containing the lowest action value at $\beta = 36$ (i.e. the model-dominated region) to represent that particular optimization design.  Fig.~\ref{fig4} shows all 16 representative paths).}
    \label{figs_appB}
\end{figure*}

\section{ACKNOWLEDGEMENTS}

The authors would like to acknowledge Tianshu Wang for providing the 1D CCSN profiles that were used in our calculations. E.~A., C.~L., L.~N., S.~R., H.~T., and Y.I. acknowledge an Institutional Support for Research and Creativity grant from New York Institute of Technology, and NSF PHY-2310066.  The research of A.~B.~B. was supported in part by the U.S. National Science Foundation Grants No. PHY-2020275, PHY-2108339 and PHY-2411495. The work of AVP was supported by the U.S. Department of Energy (DOE) under grant DE-FG02-87ER40328 at the University of Minnesota. All authors thank the good people of Kansas and Doylestown, Ohio.

\bibliography{bib_combined}

\begin{thebibliography}{69}%
\makeatletter
\providecommand \@ifxundefined [1]{%
 \@ifx{#1\undefined}
}%
\providecommand \@ifnum [1]{%
 \ifnum #1\expandafter \@firstoftwo
 \else \expandafter \@secondoftwo
 \fi
}%
\providecommand \@ifx [1]{%
 \ifx #1\expandafter \@firstoftwo
 \else \expandafter \@secondoftwo
 \fi
}%
\providecommand \natexlab [1]{#1}%
\providecommand \enquote  [1]{``#1''}%
\providecommand \bibnamefont  [1]{#1}%
\providecommand \bibfnamefont [1]{#1}%
\providecommand \citenamefont [1]{#1}%
\providecommand \href@noop [0]{\@secondoftwo}%
\providecommand \href [0]{\begingroup \@sanitize@url \@href}%
\providecommand \@href[1]{\@@startlink{#1}\@@href}%
\providecommand \@@href[1]{\endgroup#1\@@endlink}%
\providecommand \@sanitize@url [0]{\catcode `\\12\catcode `\$12\catcode
  `\&12\catcode `\#12\catcode `\^12\catcode `\_12\catcode `\%12\relax}%
\providecommand \@@startlink[1]{}%
\providecommand \@@endlink[0]{}%
\providecommand \url  [0]{\begingroup\@sanitize@url \@url }%
\providecommand \@url [1]{\endgroup\@href {#1}{\urlprefix }}%
\providecommand \urlprefix  [0]{URL }%
\providecommand \Eprint [0]{\href }%
\providecommand \doibase [0]{http://dx.doi.org/}%
\providecommand \selectlanguage [0]{\@gobble}%
\providecommand \bibinfo  [0]{\@secondoftwo}%
\providecommand \bibfield  [0]{\@secondoftwo}%
\providecommand \translation [1]{[#1]}%
\providecommand \BibitemOpen [0]{}%
\providecommand \bibitemStop [0]{}%
\providecommand \bibitemNoStop [0]{.\EOS\space}%
\providecommand \EOS [0]{\spacefactor3000\relax}%
\providecommand \BibitemShut  [1]{\csname bibitem#1\endcsname}%
\let\auto@bib@innerbib\@empty
\bibitem [{\citenamefont {{Fuller}}\ \emph {et~al.}(1992)\citenamefont
  {{Fuller}}, \citenamefont {{Mayle}}, \citenamefont {{Meyer}},\ and\
  \citenamefont {{Wilson}}}]{Fuller:1992eu}%
  \BibitemOpen
  \bibfield  {author} {\bibinfo {author} {\bibfnamefont {G.~M.}\ \bibnamefont
  {{Fuller}}}, \bibinfo {author} {\bibfnamefont {R.}~\bibnamefont {{Mayle}}},
  \bibinfo {author} {\bibfnamefont {B.~S.}\ \bibnamefont {{Meyer}}}, \ and\
  \bibinfo {author} {\bibfnamefont {J.~R.}\ \bibnamefont {{Wilson}}},\ }\href
  {\doibase 10.1086/171228} {\bibfield  {journal} {\bibinfo  {journal}
  {{Astrophysical Journal}}\ }\textbf {\bibinfo {volume} {389}},\ \bibinfo
  {pages} {517} (\bibinfo {year} {1992})}\BibitemShut {NoStop}%
\bibitem [{\citenamefont {Qian}\ \emph {et~al.}(1993)\citenamefont {Qian},
  \citenamefont {Fuller}, \citenamefont {Mathews}, \citenamefont {Mayle},
  \citenamefont {Wilson},\ and\ \citenamefont {Woosley}}]{Qian:1993dg}%
  \BibitemOpen
  \bibfield  {author} {\bibinfo {author} {\bibfnamefont {Y.-Z.}\ \bibnamefont
  {Qian}}, \bibinfo {author} {\bibfnamefont {G.~M.}\ \bibnamefont {Fuller}},
  \bibinfo {author} {\bibfnamefont {G.~J.}\ \bibnamefont {Mathews}}, \bibinfo
  {author} {\bibfnamefont {R.}~\bibnamefont {Mayle}}, \bibinfo {author}
  {\bibfnamefont {J.~R.}\ \bibnamefont {Wilson}}, \ and\ \bibinfo {author}
  {\bibfnamefont {S.~E.}\ \bibnamefont {Woosley}},\ }\href {\doibase
  10.1103/PhysRevLett.71.1965} {\bibfield  {journal} {\bibinfo  {journal}
  {Phys. Rev. Lett.}\ }\textbf {\bibinfo {volume} {71}},\ \bibinfo {pages}
  {1965} (\bibinfo {year} {1993})}\BibitemShut {NoStop}%
\bibitem [{\citenamefont {Fuller}(1993)}]{Fuller:1993ry}%
  \BibitemOpen
  \bibfield  {author} {\bibinfo {author} {\bibfnamefont {G.~M.}\ \bibnamefont
  {Fuller}},\ }\href {\doibase 10.1016/0370-1573(93)90063-J} {\bibfield
  {journal} {\bibinfo  {journal} {Phys. Rept.}\ }\textbf {\bibinfo {volume}
  {227}},\ \bibinfo {pages} {149} (\bibinfo {year} {1993})}\BibitemShut
  {NoStop}%
\bibitem [{\citenamefont {{Fuller}}\ and\ \citenamefont
  {{Meyer}}(1995)}]{Fuller:1995qy}%
  \BibitemOpen
  \bibfield  {author} {\bibinfo {author} {\bibfnamefont {G.~M.}\ \bibnamefont
  {{Fuller}}}\ and\ \bibinfo {author} {\bibfnamefont {B.~S.}\ \bibnamefont
  {{Meyer}}},\ }\href {\doibase 10.1086/176442} {\bibfield  {journal} {\bibinfo
   {journal} {{Astrophysical Journal}}\ }\textbf {\bibinfo {volume} {453}},\
  \bibinfo {pages} {792} (\bibinfo {year} {1995})}\BibitemShut {NoStop}%
\bibitem [{\citenamefont {Duan}\ \emph {et~al.}(2011)\citenamefont {Duan},
  \citenamefont {Friedland}, \citenamefont {McLaughlin},\ and\ \citenamefont
  {Surman}}]{Duan:2010af}%
  \BibitemOpen
  \bibfield  {author} {\bibinfo {author} {\bibfnamefont {H.}~\bibnamefont
  {Duan}}, \bibinfo {author} {\bibfnamefont {A.}~\bibnamefont {Friedland}},
  \bibinfo {author} {\bibfnamefont {G.}~\bibnamefont {McLaughlin}}, \ and\
  \bibinfo {author} {\bibfnamefont {R.}~\bibnamefont {Surman}},\ }\href
  {\doibase 10.1088/0954-3899/38/3/035201} {\bibfield  {journal} {\bibinfo
  {journal} {J. Phys. G}\ }\textbf {\bibinfo {volume} {38}},\ \bibinfo {pages}
  {035201} (\bibinfo {year} {2011})},\ \Eprint {http://arxiv.org/abs/1012.0532}
  {arXiv:1012.0532 [astro-ph.SR]} \BibitemShut {NoStop}%
\bibitem [{\citenamefont {Wu}\ \emph {et~al.}(2015)\citenamefont {Wu},
  \citenamefont {Qian}, \citenamefont {Martinez-Pinedo}, \citenamefont
  {Fischer},\ and\ \citenamefont {Huther}}]{Wu:2014kaa}%
  \BibitemOpen
  \bibfield  {author} {\bibinfo {author} {\bibfnamefont {M.-R.}\ \bibnamefont
  {Wu}}, \bibinfo {author} {\bibfnamefont {Y.-Z.}\ \bibnamefont {Qian}},
  \bibinfo {author} {\bibfnamefont {G.}~\bibnamefont {Martinez-Pinedo}},
  \bibinfo {author} {\bibfnamefont {T.}~\bibnamefont {Fischer}}, \ and\
  \bibinfo {author} {\bibfnamefont {L.}~\bibnamefont {Huther}},\ }\href
  {\doibase 10.1103/PhysRevD.91.065016} {\bibfield  {journal} {\bibinfo
  {journal} {Phys. Rev. D}\ }\textbf {\bibinfo {volume} {91}},\ \bibinfo
  {pages} {065016} (\bibinfo {year} {2015})},\ \Eprint
  {http://arxiv.org/abs/1412.8587} {arXiv:1412.8587 [astro-ph.HE]} \BibitemShut
  {NoStop}%
\bibitem [{\citenamefont {Wu}\ \emph {et~al.}(2016)\citenamefont {Wu},
  \citenamefont {Martinez-Pinedo},\ and\ \citenamefont {Qian}}]{Wu:2015glr}%
  \BibitemOpen
  \bibfield  {author} {\bibinfo {author} {\bibfnamefont {M.-R.}\ \bibnamefont
  {Wu}}, \bibinfo {author} {\bibfnamefont {G.}~\bibnamefont {Martinez-Pinedo}},
  \ and\ \bibinfo {author} {\bibfnamefont {Y.-Z.}\ \bibnamefont {Qian}},\
  }\href {\doibase 10.1051/epjconf/201610906005} {\bibfield  {journal}
  {\bibinfo  {journal} {EPJ Web Conf.}\ }\textbf {\bibinfo {volume} {109}},\
  \bibinfo {pages} {06005} (\bibinfo {year} {2016})},\ \Eprint
  {http://arxiv.org/abs/1512.03630} {arXiv:1512.03630 [astro-ph.HE]}
  \BibitemShut {NoStop}%
\bibitem [{\citenamefont {Sasaki}\ \emph {et~al.}(2017)\citenamefont {Sasaki},
  \citenamefont {Kajino}, \citenamefont {Takiwaki}, \citenamefont {Hayakawa},
  \citenamefont {Balantekin},\ and\ \citenamefont {Pehlivan}}]{Sasaki:2017jry}%
  \BibitemOpen
  \bibfield  {author} {\bibinfo {author} {\bibfnamefont {H.}~\bibnamefont
  {Sasaki}}, \bibinfo {author} {\bibfnamefont {T.}~\bibnamefont {Kajino}},
  \bibinfo {author} {\bibfnamefont {T.}~\bibnamefont {Takiwaki}}, \bibinfo
  {author} {\bibfnamefont {T.}~\bibnamefont {Hayakawa}}, \bibinfo {author}
  {\bibfnamefont {A.~B.}\ \bibnamefont {Balantekin}}, \ and\ \bibinfo {author}
  {\bibfnamefont {Y.}~\bibnamefont {Pehlivan}},\ }\href {\doibase
  10.1103/PhysRevD.96.043013} {\bibfield  {journal} {\bibinfo  {journal} {Phys.
  Rev. D}\ }\textbf {\bibinfo {volume} {96}},\ \bibinfo {pages} {043013}
  (\bibinfo {year} {2017})},\ \Eprint {http://arxiv.org/abs/1707.09111}
  {arXiv:1707.09111 [astro-ph.HE]} \BibitemShut {NoStop}%
\bibitem [{\citenamefont {Balantekin}(2018)}]{Balantekin:2017bau}%
  \BibitemOpen
  \bibfield  {author} {\bibinfo {author} {\bibfnamefont {A.~B.}\ \bibnamefont
  {Balantekin}},\ }\href {\doibase 10.1063/1.5030816} {\bibfield  {journal}
  {\bibinfo  {journal} {AIP Conf. Proc.}\ }\textbf {\bibinfo {volume} {1947}},\
  \bibinfo {pages} {020012} (\bibinfo {year} {2018})},\ \Eprint
  {http://arxiv.org/abs/1710.04108} {arXiv:1710.04108 [nucl-th]} \BibitemShut
  {NoStop}%
\bibitem [{\citenamefont {Xiong}\ \emph {et~al.}(2019)\citenamefont {Xiong},
  \citenamefont {Wu},\ and\ \citenamefont {Qian}}]{Xiong:2019nvw}%
  \BibitemOpen
  \bibfield  {author} {\bibinfo {author} {\bibfnamefont {Z.}~\bibnamefont
  {Xiong}}, \bibinfo {author} {\bibfnamefont {M.-R.}\ \bibnamefont {Wu}}, \
  and\ \bibinfo {author} {\bibfnamefont {Y.-Z.}\ \bibnamefont {Qian}},\ }\href
  {\doibase 10.3847/1538-4357/ab2870} {\  (\bibinfo {year} {2019}),\
  10.3847/1538-4357/ab2870},\ \Eprint {http://arxiv.org/abs/1904.09371}
  {arXiv:1904.09371 [astro-ph.HE]} \BibitemShut {NoStop}%
\bibitem [{\citenamefont {Xiong}\ \emph {et~al.}(2020)\citenamefont {Xiong},
  \citenamefont {Sieverding}, \citenamefont {Sen},\ and\ \citenamefont
  {Qian}}]{Xiong:2020ntn}%
  \BibitemOpen
  \bibfield  {author} {\bibinfo {author} {\bibfnamefont {Z.}~\bibnamefont
  {Xiong}}, \bibinfo {author} {\bibfnamefont {A.}~\bibnamefont {Sieverding}},
  \bibinfo {author} {\bibfnamefont {M.}~\bibnamefont {Sen}}, \ and\ \bibinfo
  {author} {\bibfnamefont {Y.-Z.}\ \bibnamefont {Qian}},\ }\href {\doibase
  10.3847/1538-4357/abac5e} {\bibfield  {journal} {\bibinfo  {journal}
  {Astrophys. J.}\ }\textbf {\bibinfo {volume} {900}},\ \bibinfo {pages} {144}
  (\bibinfo {year} {2020})},\ \Eprint {http://arxiv.org/abs/2006.11414}
  {arXiv:2006.11414 [astro-ph.HE]} \BibitemShut {NoStop}%
\bibitem [{\citenamefont {Duan}\ and\ \citenamefont
  {Kneller}(2009)}]{Duan:2009cd}%
  \BibitemOpen
  \bibfield  {author} {\bibinfo {author} {\bibfnamefont {H.}~\bibnamefont
  {Duan}}\ and\ \bibinfo {author} {\bibfnamefont {J.~P.}\ \bibnamefont
  {Kneller}},\ }\href {\doibase 10.1088/0954-3899/36/11/113201} {\bibfield
  {journal} {\bibinfo  {journal} {J. Phys. G}\ }\textbf {\bibinfo {volume}
  {36}},\ \bibinfo {pages} {113201} (\bibinfo {year} {2009})},\ \Eprint
  {http://arxiv.org/abs/0904.0974} {arXiv:0904.0974 [astro-ph.HE]} \BibitemShut
  {NoStop}%
\bibitem [{\citenamefont {Duan}\ \emph
  {et~al.}(2010{\natexlab{a}})\citenamefont {Duan}, \citenamefont {Fuller},\
  and\ \citenamefont {Qian}}]{Duan:2010bg}%
  \BibitemOpen
  \bibfield  {author} {\bibinfo {author} {\bibfnamefont {H.}~\bibnamefont
  {Duan}}, \bibinfo {author} {\bibfnamefont {G.~M.}\ \bibnamefont {Fuller}}, \
  and\ \bibinfo {author} {\bibfnamefont {Y.-Z.}\ \bibnamefont {Qian}},\ }\href
  {\doibase 10.1146/annurev.nucl.012809.104524} {\bibfield  {journal} {\bibinfo
   {journal} {Ann. Rev. Nucl. Part. Sci.}\ }\textbf {\bibinfo {volume} {60}},\
  \bibinfo {pages} {569} (\bibinfo {year} {2010}{\natexlab{a}})},\ \Eprint
  {http://arxiv.org/abs/1001.2799} {arXiv:1001.2799 [hep-ph]} \BibitemShut
  {NoStop}%
\bibitem [{\citenamefont {Mirizzi}\ \emph {et~al.}(2016)\citenamefont
  {Mirizzi}, \citenamefont {Tamborra}, \citenamefont {Janka}, \citenamefont
  {Saviano}, \citenamefont {Scholberg}, \citenamefont {Bollig}, \citenamefont
  {Hudepohl},\ and\ \citenamefont {Chakraborty}}]{Mirizzi:2015eza}%
  \BibitemOpen
  \bibfield  {author} {\bibinfo {author} {\bibfnamefont {A.}~\bibnamefont
  {Mirizzi}}, \bibinfo {author} {\bibfnamefont {I.}~\bibnamefont {Tamborra}},
  \bibinfo {author} {\bibfnamefont {H.-T.}\ \bibnamefont {Janka}}, \bibinfo
  {author} {\bibfnamefont {N.}~\bibnamefont {Saviano}}, \bibinfo {author}
  {\bibfnamefont {K.}~\bibnamefont {Scholberg}}, \bibinfo {author}
  {\bibfnamefont {R.}~\bibnamefont {Bollig}}, \bibinfo {author} {\bibfnamefont
  {L.}~\bibnamefont {Hudepohl}}, \ and\ \bibinfo {author} {\bibfnamefont
  {S.}~\bibnamefont {Chakraborty}},\ }\href {\doibase
  10.1393/ncr/i2016-10120-8} {\bibfield  {journal} {\bibinfo  {journal} {Riv.
  Nuovo Cim.}\ }\textbf {\bibinfo {volume} {39}},\ \bibinfo {pages} {1}
  (\bibinfo {year} {2016})},\ \Eprint {http://arxiv.org/abs/1508.00785}
  {arXiv:1508.00785 [astro-ph.HE]} \BibitemShut {NoStop}%
\bibitem [{\citenamefont {Chakraborty}\ \emph {et~al.}(2016)\citenamefont
  {Chakraborty}, \citenamefont {Hansen}, \citenamefont {Izaguirre},\ and\
  \citenamefont {Raffelt}}]{Chakraborty:2016yeg}%
  \BibitemOpen
  \bibfield  {author} {\bibinfo {author} {\bibfnamefont {S.}~\bibnamefont
  {Chakraborty}}, \bibinfo {author} {\bibfnamefont {R.}~\bibnamefont {Hansen}},
  \bibinfo {author} {\bibfnamefont {I.}~\bibnamefont {Izaguirre}}, \ and\
  \bibinfo {author} {\bibfnamefont {G.}~\bibnamefont {Raffelt}},\ }\href
  {\doibase 10.1016/j.nuclphysb.2016.02.012} {\bibfield  {journal} {\bibinfo
  {journal} {Nucl. Phys. B}\ }\textbf {\bibinfo {volume} {908}},\ \bibinfo
  {pages} {366} (\bibinfo {year} {2016})},\ \Eprint
  {http://arxiv.org/abs/1602.02766} {arXiv:1602.02766 [hep-ph]} \BibitemShut
  {NoStop}%
\bibitem [{\citenamefont {Tamborra}\ and\ \citenamefont
  {Shalgar}(2021)}]{Tamborra:2020cul}%
  \BibitemOpen
  \bibfield  {author} {\bibinfo {author} {\bibfnamefont {I.}~\bibnamefont
  {Tamborra}}\ and\ \bibinfo {author} {\bibfnamefont {S.}~\bibnamefont
  {Shalgar}},\ }\href {\doibase 10.1146/annurev-nucl-102920-050505} {\bibfield
  {journal} {\bibinfo  {journal} {Ann. Rev. Nucl. Part. Sci.}\ }\textbf
  {\bibinfo {volume} {71}},\ \bibinfo {pages} {165} (\bibinfo {year} {2021})},\
  \Eprint {http://arxiv.org/abs/2011.01948} {arXiv:2011.01948 [astro-ph.HE]}
  \BibitemShut {NoStop}%
\bibitem [{\citenamefont {Richers}\ and\ \citenamefont
  {Sen}(2022)}]{Richers:2022zug}%
  \BibitemOpen
  \bibfield  {author} {\bibinfo {author} {\bibfnamefont {S.}~\bibnamefont
  {Richers}}\ and\ \bibinfo {author} {\bibfnamefont {M.}~\bibnamefont {Sen}},\
  }\enquote {\bibinfo {title} {{Fast Flavor Transformations}},}\ in\ \href
  {\doibase 10.1007/978-981-15-8818-1_125-1} {\emph {\bibinfo {booktitle}
  {{Handbook of Nuclear Physics}}}},\ \bibinfo {editor} {edited by\ \bibinfo
  {editor} {\bibfnamefont {I.}~\bibnamefont {Tanihata}}, \bibinfo {editor}
  {\bibfnamefont {H.}~\bibnamefont {Toki}}, \ and\ \bibinfo {editor}
  {\bibfnamefont {T.}~\bibnamefont {Kajino}}}\ (\bibinfo {year} {2022})\ pp.\
  \bibinfo {pages} {1--17},\ \Eprint {http://arxiv.org/abs/2207.03561}
  {arXiv:2207.03561 [astro-ph.HE]} \BibitemShut {NoStop}%
\bibitem [{\citenamefont {Duan}\ \emph {et~al.}(2008)\citenamefont {Duan},
  \citenamefont {Fuller},\ and\ \citenamefont {Carlson}}]{duan2008simulating}%
  \BibitemOpen
  \bibfield  {author} {\bibinfo {author} {\bibfnamefont {H.}~\bibnamefont
  {Duan}}, \bibinfo {author} {\bibfnamefont {G.~M.}\ \bibnamefont {Fuller}}, \
  and\ \bibinfo {author} {\bibfnamefont {J.}~\bibnamefont {Carlson}},\
  }\href@noop {} {\bibfield  {journal} {\bibinfo  {journal} {Computational
  Science \& Discovery}\ }\textbf {\bibinfo {volume} {1}},\ \bibinfo {pages}
  {015007} (\bibinfo {year} {2008})}\BibitemShut {NoStop}%
\bibitem [{\citenamefont {Richers}\ \emph {et~al.}(2019)\citenamefont
  {Richers}, \citenamefont {McLaughlin}, \citenamefont {Kneller},\ and\
  \citenamefont {Vlasenko}}]{richers2019neutrino}%
  \BibitemOpen
  \bibfield  {author} {\bibinfo {author} {\bibfnamefont {S.~A.}\ \bibnamefont
  {Richers}}, \bibinfo {author} {\bibfnamefont {G.~C.}\ \bibnamefont
  {McLaughlin}}, \bibinfo {author} {\bibfnamefont {J.~P.}\ \bibnamefont
  {Kneller}}, \ and\ \bibinfo {author} {\bibfnamefont {A.}~\bibnamefont
  {Vlasenko}},\ }\href@noop {} {\bibfield  {journal} {\bibinfo  {journal}
  {Physical Review D}\ }\textbf {\bibinfo {volume} {99}},\ \bibinfo {pages}
  {123014} (\bibinfo {year} {2019})}\BibitemShut {NoStop}%
\bibitem [{\citenamefont {Richers}\ \emph {et~al.}(2021)\citenamefont
  {Richers}, \citenamefont {Willcox}, \citenamefont {Ford},\ and\ \citenamefont
  {Myers}}]{Richers:2021nbx}%
  \BibitemOpen
  \bibfield  {author} {\bibinfo {author} {\bibfnamefont {S.}~\bibnamefont
  {Richers}}, \bibinfo {author} {\bibfnamefont {D.~E.}\ \bibnamefont
  {Willcox}}, \bibinfo {author} {\bibfnamefont {N.~M.}\ \bibnamefont {Ford}}, \
  and\ \bibinfo {author} {\bibfnamefont {A.}~\bibnamefont {Myers}},\ }\href
  {\doibase 10.1103/PhysRevD.103.083013} {\bibfield  {journal} {\bibinfo
  {journal} {Phys. Rev. D}\ }\textbf {\bibinfo {volume} {103}},\ \bibinfo
  {pages} {083013} (\bibinfo {year} {2021})},\ \Eprint
  {http://arxiv.org/abs/2101.02745} {arXiv:2101.02745 [astro-ph.HE]}
  \BibitemShut {NoStop}%
\bibitem [{\citenamefont {Bhattacharyya}\ and\ \citenamefont
  {Dasgupta}(2021)}]{Bhattacharyya:2020jpj}%
  \BibitemOpen
  \bibfield  {author} {\bibinfo {author} {\bibfnamefont {S.}~\bibnamefont
  {Bhattacharyya}}\ and\ \bibinfo {author} {\bibfnamefont {B.}~\bibnamefont
  {Dasgupta}},\ }\href {\doibase 10.1103/PhysRevLett.126.061302} {\bibfield
  {journal} {\bibinfo  {journal} {Phys. Rev. Lett.}\ }\textbf {\bibinfo
  {volume} {126}},\ \bibinfo {pages} {061302} (\bibinfo {year} {2021})},\
  \Eprint {http://arxiv.org/abs/2009.03337} {arXiv:2009.03337 [hep-ph]}
  \BibitemShut {NoStop}%
\bibitem [{\citenamefont {Zaizen}\ and\ \citenamefont
  {Morinaga}(2021)}]{Zaizen:2021wwl}%
  \BibitemOpen
  \bibfield  {author} {\bibinfo {author} {\bibfnamefont {M.}~\bibnamefont
  {Zaizen}}\ and\ \bibinfo {author} {\bibfnamefont {T.}~\bibnamefont
  {Morinaga}},\ }\href {\doibase 10.1103/PhysRevD.104.083035} {\bibfield
  {journal} {\bibinfo  {journal} {Phys. Rev. D}\ }\textbf {\bibinfo {volume}
  {104}},\ \bibinfo {pages} {083035} (\bibinfo {year} {2021})},\ \Eprint
  {http://arxiv.org/abs/2104.10532} {arXiv:2104.10532 [hep-ph]} \BibitemShut
  {NoStop}%
\bibitem [{\citenamefont {Martin}\ \emph {et~al.}(2019)\citenamefont {Martin},
  \citenamefont {Abbar},\ and\ \citenamefont {Duan}}]{Martin:2019kgi}%
  \BibitemOpen
  \bibfield  {author} {\bibinfo {author} {\bibfnamefont {J.~D.}\ \bibnamefont
  {Martin}}, \bibinfo {author} {\bibfnamefont {S.}~\bibnamefont {Abbar}}, \
  and\ \bibinfo {author} {\bibfnamefont {H.}~\bibnamefont {Duan}},\ }\href
  {\doibase 10.1103/PhysRevD.100.023016} {\bibfield  {journal} {\bibinfo
  {journal} {Phys. Rev. D}\ }\textbf {\bibinfo {volume} {100}},\ \bibinfo
  {pages} {023016} (\bibinfo {year} {2019})},\ \Eprint
  {http://arxiv.org/abs/1904.08877} {arXiv:1904.08877 [hep-ph]} \BibitemShut
  {NoStop}%
\bibitem [{\citenamefont {George}\ \emph {et~al.}(2023)\citenamefont {George},
  \citenamefont {Lin}, \citenamefont {Wu}, \citenamefont {Liu},\ and\
  \citenamefont {Xiong}}]{George:2022lwg}%
  \BibitemOpen
  \bibfield  {author} {\bibinfo {author} {\bibfnamefont {M.}~\bibnamefont
  {George}}, \bibinfo {author} {\bibfnamefont {C.-Y.}\ \bibnamefont {Lin}},
  \bibinfo {author} {\bibfnamefont {M.-R.}\ \bibnamefont {Wu}}, \bibinfo
  {author} {\bibfnamefont {T.~G.}\ \bibnamefont {Liu}}, \ and\ \bibinfo
  {author} {\bibfnamefont {Z.}~\bibnamefont {Xiong}},\ }\href {\doibase
  10.1016/j.cpc.2022.108588} {\bibfield  {journal} {\bibinfo  {journal}
  {Comput. Phys. Commun.}\ }\textbf {\bibinfo {volume} {283}},\ \bibinfo
  {pages} {108588} (\bibinfo {year} {2023})},\ \Eprint
  {http://arxiv.org/abs/2203.12866} {arXiv:2203.12866 [hep-ph]} \BibitemShut
  {NoStop}%
\bibitem [{\citenamefont {Richers}\ \emph {et~al.}(2022)\citenamefont
  {Richers}, \citenamefont {Duan}, \citenamefont {Wu}, \citenamefont
  {Bhattacharyya}, \citenamefont {Zaizen}, \citenamefont {George},
  \citenamefont {Lin},\ and\ \citenamefont {Xiong}}]{Richers:2022bkd}%
  \BibitemOpen
  \bibfield  {author} {\bibinfo {author} {\bibfnamefont {S.}~\bibnamefont
  {Richers}}, \bibinfo {author} {\bibfnamefont {H.}~\bibnamefont {Duan}},
  \bibinfo {author} {\bibfnamefont {M.-R.}\ \bibnamefont {Wu}}, \bibinfo
  {author} {\bibfnamefont {S.}~\bibnamefont {Bhattacharyya}}, \bibinfo {author}
  {\bibfnamefont {M.}~\bibnamefont {Zaizen}}, \bibinfo {author} {\bibfnamefont
  {M.}~\bibnamefont {George}}, \bibinfo {author} {\bibfnamefont {C.-Y.}\
  \bibnamefont {Lin}}, \ and\ \bibinfo {author} {\bibfnamefont
  {Z.}~\bibnamefont {Xiong}},\ }\href {\doibase 10.1103/PhysRevD.106.043011}
  {\bibfield  {journal} {\bibinfo  {journal} {Phys. Rev. D}\ }\textbf {\bibinfo
  {volume} {106}},\ \bibinfo {pages} {043011} (\bibinfo {year} {2022})},\
  \Eprint {http://arxiv.org/abs/2205.06282} {arXiv:2205.06282 [astro-ph.HE]}
  \BibitemShut {NoStop}%
\bibitem [{\citenamefont {Cherry}\ \emph {et~al.}(2012)\citenamefont {Cherry},
  \citenamefont {Carlson}, \citenamefont {Friedland}, \citenamefont {Fuller},\
  and\ \citenamefont {Vlasenko}}]{cherry2012neutrino}%
  \BibitemOpen
  \bibfield  {author} {\bibinfo {author} {\bibfnamefont {J.~F.}\ \bibnamefont
  {Cherry}}, \bibinfo {author} {\bibfnamefont {J.}~\bibnamefont {Carlson}},
  \bibinfo {author} {\bibfnamefont {A.}~\bibnamefont {Friedland}}, \bibinfo
  {author} {\bibfnamefont {G.~M.}\ \bibnamefont {Fuller}}, \ and\ \bibinfo
  {author} {\bibfnamefont {A.}~\bibnamefont {Vlasenko}},\ }\href@noop {}
  {\bibfield  {journal} {\bibinfo  {journal} {Physical review letters}\
  }\textbf {\bibinfo {volume} {108}},\ \bibinfo {pages} {261104} (\bibinfo
  {year} {2012})}\BibitemShut {NoStop}%
\bibitem [{\citenamefont {Cherry}\ \emph {et~al.}(2013)\citenamefont {Cherry},
  \citenamefont {Carlson}, \citenamefont {Friedland}, \citenamefont {Fuller},\
  and\ \citenamefont {Vlasenko}}]{Cherry:2013mv}%
  \BibitemOpen
  \bibfield  {author} {\bibinfo {author} {\bibfnamefont {J.~F.}\ \bibnamefont
  {Cherry}}, \bibinfo {author} {\bibfnamefont {J.}~\bibnamefont {Carlson}},
  \bibinfo {author} {\bibfnamefont {A.}~\bibnamefont {Friedland}}, \bibinfo
  {author} {\bibfnamefont {G.~M.}\ \bibnamefont {Fuller}}, \ and\ \bibinfo
  {author} {\bibfnamefont {A.}~\bibnamefont {Vlasenko}},\ }\href {\doibase
  10.1103/PhysRevD.87.085037} {\bibfield  {journal} {\bibinfo  {journal} {Phys.
  Rev. D}\ }\textbf {\bibinfo {volume} {87}},\ \bibinfo {pages} {085037}
  (\bibinfo {year} {2013})},\ \Eprint {http://arxiv.org/abs/1302.1159}
  {arXiv:1302.1159 [astro-ph.HE]} \BibitemShut {NoStop}%
\bibitem [{\citenamefont {Zaizen}\ \emph {et~al.}(2020)\citenamefont {Zaizen},
  \citenamefont {Cherry}, \citenamefont {Takiwaki}, \citenamefont {Horiuchi},
  \citenamefont {Kotake}, \citenamefont {Umeda},\ and\ \citenamefont
  {Yoshida}}]{Zaizen:2019ufj}%
  \BibitemOpen
  \bibfield  {author} {\bibinfo {author} {\bibfnamefont {M.}~\bibnamefont
  {Zaizen}}, \bibinfo {author} {\bibfnamefont {J.~F.}\ \bibnamefont {Cherry}},
  \bibinfo {author} {\bibfnamefont {T.}~\bibnamefont {Takiwaki}}, \bibinfo
  {author} {\bibfnamefont {S.}~\bibnamefont {Horiuchi}}, \bibinfo {author}
  {\bibfnamefont {K.}~\bibnamefont {Kotake}}, \bibinfo {author} {\bibfnamefont
  {H.}~\bibnamefont {Umeda}}, \ and\ \bibinfo {author} {\bibfnamefont
  {T.}~\bibnamefont {Yoshida}},\ }\href {\doibase
  10.1088/1475-7516/2020/06/011} {\bibfield  {journal} {\bibinfo  {journal}
  {JCAP}\ }\textbf {\bibinfo {volume} {06}},\ \bibinfo {pages} {011} (\bibinfo
  {year} {2020})},\ \Eprint {http://arxiv.org/abs/1908.10594} {arXiv:1908.10594
  [astro-ph.HE]} \BibitemShut {NoStop}%
\bibitem [{\citenamefont {Cherry}\ \emph {et~al.}(2019)\citenamefont {Cherry},
  \citenamefont {Fuller}, \citenamefont {Horiuchi}, \citenamefont {Kotake},
  \citenamefont {Takiwaki},\ and\ \citenamefont {Fischer}}]{Cherry:2019vkv}%
  \BibitemOpen
  \bibfield  {author} {\bibinfo {author} {\bibfnamefont {J.~F.}\ \bibnamefont
  {Cherry}}, \bibinfo {author} {\bibfnamefont {G.~M.}\ \bibnamefont {Fuller}},
  \bibinfo {author} {\bibfnamefont {S.}~\bibnamefont {Horiuchi}}, \bibinfo
  {author} {\bibfnamefont {K.}~\bibnamefont {Kotake}}, \bibinfo {author}
  {\bibfnamefont {T.}~\bibnamefont {Takiwaki}}, \ and\ \bibinfo {author}
  {\bibfnamefont {T.}~\bibnamefont {Fischer}},\ }\href@noop {} {\  (\bibinfo
  {year} {2019})},\ \Eprint {http://arxiv.org/abs/1912.11489} {arXiv:1912.11489
  [astro-ph.HE]} \BibitemShut {NoStop}%
\bibitem [{\citenamefont {Cirigliano}\ \emph {et~al.}(2018)\citenamefont
  {Cirigliano}, \citenamefont {Paris},\ and\ \citenamefont
  {Shalgar}}]{cirigliano2018collective}%
  \BibitemOpen
  \bibfield  {author} {\bibinfo {author} {\bibfnamefont {V.}~\bibnamefont
  {Cirigliano}}, \bibinfo {author} {\bibfnamefont {M.}~\bibnamefont {Paris}}, \
  and\ \bibinfo {author} {\bibfnamefont {S.}~\bibnamefont {Shalgar}},\
  }\href@noop {} {\bibfield  {journal} {\bibinfo  {journal} {Journal of
  Cosmology and Astroparticle Physics}\ }\textbf {\bibinfo {volume} {2018}},\
  \bibinfo {pages} {019} (\bibinfo {year} {2018})}\BibitemShut {NoStop}%
\bibitem [{\citenamefont {Kimura}(2002)}]{kimura2002numerical}%
  \BibitemOpen
  \bibfield  {author} {\bibinfo {author} {\bibfnamefont {R.}~\bibnamefont
  {Kimura}},\ }\href@noop {} {\bibfield  {journal} {\bibinfo  {journal}
  {Journal of Wind Engineering and Industrial Aerodynamics}\ }\textbf {\bibinfo
  {volume} {90}},\ \bibinfo {pages} {1403} (\bibinfo {year}
  {2002})}\BibitemShut {NoStop}%
\bibitem [{\citenamefont {Kalnay}(2003)}]{kalnay2003atmospheric}%
  \BibitemOpen
  \bibfield  {author} {\bibinfo {author} {\bibfnamefont {E.}~\bibnamefont
  {Kalnay}},\ }\href@noop {} {\emph {\bibinfo {title} {Atmospheric modeling,
  data assimilation and predictability}}}\ (\bibinfo  {publisher} {Cambridge
  university press},\ \bibinfo {year} {2003})\BibitemShut {NoStop}%
\bibitem [{\citenamefont {Evensen}(2009)}]{evensen2009data}%
  \BibitemOpen
  \bibfield  {author} {\bibinfo {author} {\bibfnamefont {G.}~\bibnamefont
  {Evensen}},\ }\href@noop {} {\emph {\bibinfo {title} {Data assimilation: the
  ensemble Kalman filter}}}\ (\bibinfo  {publisher} {Springer Science \&
  Business Media},\ \bibinfo {year} {2009})\BibitemShut {NoStop}%
\bibitem [{\citenamefont {Betts}(2010)}]{betts2010practical}%
  \BibitemOpen
  \bibfield  {author} {\bibinfo {author} {\bibfnamefont {J.~T.}\ \bibnamefont
  {Betts}},\ }\href@noop {} {\emph {\bibinfo {title} {Practical methods for
  optimal control and estimation using nonlinear programming}}},\ Vol.~\bibinfo
  {volume} {19}\ (\bibinfo  {publisher} {Siam},\ \bibinfo {year}
  {2010})\BibitemShut {NoStop}%
\bibitem [{\citenamefont {Whartenby}\ \emph {et~al.}(2013)\citenamefont
  {Whartenby}, \citenamefont {Quinn},\ and\ \citenamefont
  {Abarbanel}}]{whartenby2013number}%
  \BibitemOpen
  \bibfield  {author} {\bibinfo {author} {\bibfnamefont {W.~G.}\ \bibnamefont
  {Whartenby}}, \bibinfo {author} {\bibfnamefont {J.~C.}\ \bibnamefont
  {Quinn}}, \ and\ \bibinfo {author} {\bibfnamefont {H.~D.}\ \bibnamefont
  {Abarbanel}},\ }\href@noop {} {\bibfield  {journal} {\bibinfo  {journal}
  {Monthly Weather Review}\ }\textbf {\bibinfo {volume} {141}},\ \bibinfo
  {pages} {2502} (\bibinfo {year} {2013})}\BibitemShut {NoStop}%
\bibitem [{\citenamefont {An}\ \emph {et~al.}(2017)\citenamefont {An},
  \citenamefont {Rey}, \citenamefont {Ye},\ and\ \citenamefont
  {Abarbanel}}]{an2017estimating}%
  \BibitemOpen
  \bibfield  {author} {\bibinfo {author} {\bibfnamefont {Z.}~\bibnamefont
  {An}}, \bibinfo {author} {\bibfnamefont {D.}~\bibnamefont {Rey}}, \bibinfo
  {author} {\bibfnamefont {J.}~\bibnamefont {Ye}}, \ and\ \bibinfo {author}
  {\bibfnamefont {H.~D.}\ \bibnamefont {Abarbanel}},\ }\href@noop {} {\bibfield
   {journal} {\bibinfo  {journal} {Nonlinear Processes in Geophysics (Online)}\
  }\textbf {\bibinfo {volume} {24}} (\bibinfo {year} {2017})}\BibitemShut
  {NoStop}%
\bibitem [{\citenamefont {Schiff}(2009)}]{schiff2009kalman}%
  \BibitemOpen
  \bibfield  {author} {\bibinfo {author} {\bibfnamefont {S.~J.}\ \bibnamefont
  {Schiff}},\ }in\ \href@noop {} {\emph {\bibinfo {booktitle} {2009 Annual
  International Conference of the IEEE Engineering in Medicine and Biology
  Society}}}\ (\bibinfo {organization} {IEEE},\ \bibinfo {year} {2009})\ pp.\
  \bibinfo {pages} {3318--3321}\BibitemShut {NoStop}%
\bibitem [{\citenamefont {Toth}\ \emph {et~al.}(2011)\citenamefont {Toth},
  \citenamefont {Kostuk}, \citenamefont {Meliza}, \citenamefont {Margoliash},\
  and\ \citenamefont {Abarbanel}}]{toth2011dynamical}%
  \BibitemOpen
  \bibfield  {author} {\bibinfo {author} {\bibfnamefont {B.~A.}\ \bibnamefont
  {Toth}}, \bibinfo {author} {\bibfnamefont {M.}~\bibnamefont {Kostuk}},
  \bibinfo {author} {\bibfnamefont {C.~D.}\ \bibnamefont {Meliza}}, \bibinfo
  {author} {\bibfnamefont {D.}~\bibnamefont {Margoliash}}, \ and\ \bibinfo
  {author} {\bibfnamefont {H.~D.}\ \bibnamefont {Abarbanel}},\ }\href@noop {}
  {\bibfield  {journal} {\bibinfo  {journal} {Biological cybernetics}\ }\textbf
  {\bibinfo {volume} {105}},\ \bibinfo {pages} {217} (\bibinfo {year}
  {2011})}\BibitemShut {NoStop}%
\bibitem [{\citenamefont {Kostuk}\ \emph {et~al.}(2012)\citenamefont {Kostuk},
  \citenamefont {Toth}, \citenamefont {Meliza}, \citenamefont {Margoliash},\
  and\ \citenamefont {Abarbanel}}]{kostuk2012dynamical}%
  \BibitemOpen
  \bibfield  {author} {\bibinfo {author} {\bibfnamefont {M.}~\bibnamefont
  {Kostuk}}, \bibinfo {author} {\bibfnamefont {B.~A.}\ \bibnamefont {Toth}},
  \bibinfo {author} {\bibfnamefont {C.~D.}\ \bibnamefont {Meliza}}, \bibinfo
  {author} {\bibfnamefont {D.}~\bibnamefont {Margoliash}}, \ and\ \bibinfo
  {author} {\bibfnamefont {H.~D.}\ \bibnamefont {Abarbanel}},\ }\href@noop {}
  {\bibfield  {journal} {\bibinfo  {journal} {Biological cybernetics}\ }\textbf
  {\bibinfo {volume} {106}},\ \bibinfo {pages} {155} (\bibinfo {year}
  {2012})}\BibitemShut {NoStop}%
\bibitem [{\citenamefont {Hamilton}\ \emph {et~al.}(2013)\citenamefont
  {Hamilton}, \citenamefont {Berry}, \citenamefont {Peixoto},\ and\
  \citenamefont {Sauer}}]{hamilton2013real}%
  \BibitemOpen
  \bibfield  {author} {\bibinfo {author} {\bibfnamefont {F.}~\bibnamefont
  {Hamilton}}, \bibinfo {author} {\bibfnamefont {T.}~\bibnamefont {Berry}},
  \bibinfo {author} {\bibfnamefont {N.}~\bibnamefont {Peixoto}}, \ and\
  \bibinfo {author} {\bibfnamefont {T.}~\bibnamefont {Sauer}},\ }\href@noop {}
  {\bibfield  {journal} {\bibinfo  {journal} {Physical Review E}\ }\textbf
  {\bibinfo {volume} {88}},\ \bibinfo {pages} {052715} (\bibinfo {year}
  {2013})}\BibitemShut {NoStop}%
\bibitem [{\citenamefont {Meliza}\ \emph {et~al.}(2014)\citenamefont {Meliza},
  \citenamefont {Kostuk}, \citenamefont {Huang}, \citenamefont {Nogaret},
  \citenamefont {Margoliash},\ and\ \citenamefont
  {Abarbanel}}]{meliza2014estimating}%
  \BibitemOpen
  \bibfield  {author} {\bibinfo {author} {\bibfnamefont {C.~D.}\ \bibnamefont
  {Meliza}}, \bibinfo {author} {\bibfnamefont {M.}~\bibnamefont {Kostuk}},
  \bibinfo {author} {\bibfnamefont {H.}~\bibnamefont {Huang}}, \bibinfo
  {author} {\bibfnamefont {A.}~\bibnamefont {Nogaret}}, \bibinfo {author}
  {\bibfnamefont {D.}~\bibnamefont {Margoliash}}, \ and\ \bibinfo {author}
  {\bibfnamefont {H.~D.}\ \bibnamefont {Abarbanel}},\ }\href@noop {} {\bibfield
   {journal} {\bibinfo  {journal} {Biological cybernetics}\ }\textbf {\bibinfo
  {volume} {108}},\ \bibinfo {pages} {495} (\bibinfo {year}
  {2014})}\BibitemShut {NoStop}%
\bibitem [{\citenamefont {Nogaret}\ \emph {et~al.}(2016)\citenamefont
  {Nogaret}, \citenamefont {Meliza}, \citenamefont {Margoliash},\ and\
  \citenamefont {Abarbanel}}]{nogaret2016automatic}%
  \BibitemOpen
  \bibfield  {author} {\bibinfo {author} {\bibfnamefont {A.}~\bibnamefont
  {Nogaret}}, \bibinfo {author} {\bibfnamefont {C.~D.}\ \bibnamefont {Meliza}},
  \bibinfo {author} {\bibfnamefont {D.}~\bibnamefont {Margoliash}}, \ and\
  \bibinfo {author} {\bibfnamefont {H.~D.}\ \bibnamefont {Abarbanel}},\
  }\href@noop {} {\bibfield  {journal} {\bibinfo  {journal} {Scientific
  reports}\ }\textbf {\bibinfo {volume} {6}},\ \bibinfo {pages} {1} (\bibinfo
  {year} {2016})}\BibitemShut {NoStop}%
\bibitem [{\citenamefont {Armstrong}(2020)}]{armstrong2020statistical}%
  \BibitemOpen
  \bibfield  {author} {\bibinfo {author} {\bibfnamefont {E.}~\bibnamefont
  {Armstrong}},\ }\href@noop {} {\bibfield  {journal} {\bibinfo  {journal}
  {Physical Review E}\ }\textbf {\bibinfo {volume} {101}},\ \bibinfo {pages}
  {012415} (\bibinfo {year} {2020})}\BibitemShut {NoStop}%
\bibitem [{\citenamefont {Armstrong}\ \emph {et~al.}(2017)\citenamefont
  {Armstrong}, \citenamefont {Patwardhan}, \citenamefont {Johns}, \citenamefont
  {Kishimoto}, \citenamefont {Abarbanel},\ and\ \citenamefont
  {Fuller}}]{armstrong2017optimization}%
  \BibitemOpen
  \bibfield  {author} {\bibinfo {author} {\bibfnamefont {E.}~\bibnamefont
  {Armstrong}}, \bibinfo {author} {\bibfnamefont {A.~V.}\ \bibnamefont
  {Patwardhan}}, \bibinfo {author} {\bibfnamefont {L.}~\bibnamefont {Johns}},
  \bibinfo {author} {\bibfnamefont {C.~T.}\ \bibnamefont {Kishimoto}}, \bibinfo
  {author} {\bibfnamefont {H.~D.}\ \bibnamefont {Abarbanel}}, \ and\ \bibinfo
  {author} {\bibfnamefont {G.~M.}\ \bibnamefont {Fuller}},\ }\href@noop {}
  {\bibfield  {journal} {\bibinfo  {journal} {Physical Review D}\ }\textbf
  {\bibinfo {volume} {96}},\ \bibinfo {pages} {083008} (\bibinfo {year}
  {2017})}\BibitemShut {NoStop}%
\bibitem [{\citenamefont {Armstrong}\ \emph {et~al.}(2020)\citenamefont
  {Armstrong}, \citenamefont {Patwardhan}, \citenamefont {Rrapaj},
  \citenamefont {Ardizi},\ and\ \citenamefont
  {Fuller}}]{armstrong2020inference}%
  \BibitemOpen
  \bibfield  {author} {\bibinfo {author} {\bibfnamefont {E.}~\bibnamefont
  {Armstrong}}, \bibinfo {author} {\bibfnamefont {A.~V.}\ \bibnamefont
  {Patwardhan}}, \bibinfo {author} {\bibfnamefont {E.}~\bibnamefont {Rrapaj}},
  \bibinfo {author} {\bibfnamefont {S.~F.}\ \bibnamefont {Ardizi}}, \ and\
  \bibinfo {author} {\bibfnamefont {G.~M.}\ \bibnamefont {Fuller}},\
  }\href@noop {} {\bibfield  {journal} {\bibinfo  {journal} {Physical Review
  D}\ }\textbf {\bibinfo {volume} {102}},\ \bibinfo {pages} {043013} (\bibinfo
  {year} {2020})}\BibitemShut {NoStop}%
\bibitem [{\citenamefont {Rrapaj}\ \emph {et~al.}(2021)\citenamefont {Rrapaj},
  \citenamefont {Patwardhan}, \citenamefont {Armstrong},\ and\ \citenamefont
  {Fuller}}]{rrapaj2021inference}%
  \BibitemOpen
  \bibfield  {author} {\bibinfo {author} {\bibfnamefont {E.}~\bibnamefont
  {Rrapaj}}, \bibinfo {author} {\bibfnamefont {A.~V.}\ \bibnamefont
  {Patwardhan}}, \bibinfo {author} {\bibfnamefont {E.}~\bibnamefont
  {Armstrong}}, \ and\ \bibinfo {author} {\bibfnamefont {G.~M.}\ \bibnamefont
  {Fuller}},\ }\href@noop {} {\bibfield  {journal} {\bibinfo  {journal}
  {Physical Review D}\ }\textbf {\bibinfo {volume} {103}},\ \bibinfo {pages}
  {043006} (\bibinfo {year} {2021})}\BibitemShut {NoStop}%
\bibitem [{\citenamefont {Armstrong}(2021)}]{armstrong2021inference}%
  \BibitemOpen
  \bibfield  {author} {\bibinfo {author} {\bibfnamefont {E.}~\bibnamefont
  {Armstrong}},\ }\href@noop {} {\bibfield  {journal} {\bibinfo  {journal}
  {arXiv preprint arXiv:2111.07412}\ } (\bibinfo {year} {2021})}\BibitemShut
  {NoStop}%
\bibitem [{\citenamefont {Armstrong}(2022)}]{armstrong2022inferenceA}%
  \BibitemOpen
  \bibfield  {author} {\bibinfo {author} {\bibfnamefont {E.}~\bibnamefont
  {Armstrong}},\ }\href@noop {} {\bibfield  {journal} {\bibinfo  {journal}
  {Physical Review D}\ }\textbf {\bibinfo {volume} {105}},\ \bibinfo {pages}
  {083012} (\bibinfo {year} {2022})}\BibitemShut {NoStop}%
\bibitem [{\citenamefont {Armstrong}\ \emph {et~al.}(2022)\citenamefont
  {Armstrong}, \citenamefont {Patwardhan}, \citenamefont {Ahmetaj},
  \citenamefont {Sanchez}, \citenamefont {Miskiewicz}, \citenamefont
  {Ibrahim},\ and\ \citenamefont {Singh}}]{armstrong2022inferenceB}%
  \BibitemOpen
  \bibfield  {author} {\bibinfo {author} {\bibfnamefont {E.}~\bibnamefont
  {Armstrong}}, \bibinfo {author} {\bibfnamefont {A.~V.}\ \bibnamefont
  {Patwardhan}}, \bibinfo {author} {\bibfnamefont {A.}~\bibnamefont {Ahmetaj}},
  \bibinfo {author} {\bibfnamefont {M.~M.}\ \bibnamefont {Sanchez}}, \bibinfo
  {author} {\bibfnamefont {S.}~\bibnamefont {Miskiewicz}}, \bibinfo {author}
  {\bibfnamefont {M.}~\bibnamefont {Ibrahim}}, \ and\ \bibinfo {author}
  {\bibfnamefont {I.}~\bibnamefont {Singh}},\ }\href@noop {} {\bibfield
  {journal} {\bibinfo  {journal} {Physical Review D}\ }\textbf {\bibinfo
  {volume} {105}},\ \bibinfo {pages} {103003} (\bibinfo {year}
  {2022})}\BibitemShut {NoStop}%
\bibitem [{\citenamefont {Laber-Smith}\ \emph {et~al.}(2023)\citenamefont
  {Laber-Smith}, \citenamefont {Ahmetaj}, \citenamefont {Armstrong},
  \citenamefont {Balantekin}, \citenamefont {Patwardhan}, \citenamefont
  {Sanchez},\ and\ \citenamefont {Wong}}]{laber2023inference}%
  \BibitemOpen
  \bibfield  {author} {\bibinfo {author} {\bibfnamefont {C.}~\bibnamefont
  {Laber-Smith}}, \bibinfo {author} {\bibfnamefont {A.}~\bibnamefont
  {Ahmetaj}}, \bibinfo {author} {\bibfnamefont {E.}~\bibnamefont {Armstrong}},
  \bibinfo {author} {\bibfnamefont {A.~B.}\ \bibnamefont {Balantekin}},
  \bibinfo {author} {\bibfnamefont {A.~V.}\ \bibnamefont {Patwardhan}},
  \bibinfo {author} {\bibfnamefont {M.~M.}\ \bibnamefont {Sanchez}}, \ and\
  \bibinfo {author} {\bibfnamefont {S.}~\bibnamefont {Wong}},\ }\href@noop {}
  {\bibfield  {journal} {\bibinfo  {journal} {Physical Review D}\ }\textbf
  {\bibinfo {volume} {107}},\ \bibinfo {pages} {023013} (\bibinfo {year}
  {2023})}\BibitemShut {NoStop}%
\bibitem [{\citenamefont {Radice}\ \emph {et~al.}(2017)\citenamefont {Radice},
  \citenamefont {Burrows}, \citenamefont {Vartanyan}, \citenamefont {Skinner},\
  and\ \citenamefont {Dolence}}]{Radice:2017ykv}%
  \BibitemOpen
  \bibfield  {author} {\bibinfo {author} {\bibfnamefont {D.}~\bibnamefont
  {Radice}}, \bibinfo {author} {\bibfnamefont {A.}~\bibnamefont {Burrows}},
  \bibinfo {author} {\bibfnamefont {D.}~\bibnamefont {Vartanyan}}, \bibinfo
  {author} {\bibfnamefont {M.~A.}\ \bibnamefont {Skinner}}, \ and\ \bibinfo
  {author} {\bibfnamefont {J.~C.}\ \bibnamefont {Dolence}},\ }\href {\doibase
  10.3847/1538-4357/aa92c5} {\bibfield  {journal} {\bibinfo  {journal}
  {Astrophys. J.}\ }\textbf {\bibinfo {volume} {850}},\ \bibinfo {pages} {43}
  (\bibinfo {year} {2017})},\ \Eprint {http://arxiv.org/abs/1702.03927}
  {arXiv:1702.03927 [astro-ph.HE]} \BibitemShut {NoStop}%
\bibitem [{\citenamefont {Wang}\ and\ \citenamefont
  {Burrows}(2024)}]{Wang:2024tbv}%
  \BibitemOpen
  \bibfield  {author} {\bibinfo {author} {\bibfnamefont {T.}~\bibnamefont
  {Wang}}\ and\ \bibinfo {author} {\bibfnamefont {A.}~\bibnamefont {Burrows}},\
  }\href {\doibase 10.3847/1538-4357/ad5009} {\bibfield  {journal} {\bibinfo
  {journal} {Astrophys. J.}\ }\textbf {\bibinfo {volume} {969}},\ \bibinfo
  {pages} {74} (\bibinfo {year} {2024})},\ \Eprint
  {http://arxiv.org/abs/2405.06024} {arXiv:2405.06024 [astro-ph.HE]}
  \BibitemShut {NoStop}%
\bibitem [{\citenamefont {Nagakura}\ \emph {et~al.}(2021)\citenamefont
  {Nagakura}, \citenamefont {Burrows}, \citenamefont {Vartanyan},\ and\
  \citenamefont {Radice}}]{nagakura2021core}%
  \BibitemOpen
  \bibfield  {author} {\bibinfo {author} {\bibfnamefont {H.}~\bibnamefont
  {Nagakura}}, \bibinfo {author} {\bibfnamefont {A.}~\bibnamefont {Burrows}},
  \bibinfo {author} {\bibfnamefont {D.}~\bibnamefont {Vartanyan}}, \ and\
  \bibinfo {author} {\bibfnamefont {D.}~\bibnamefont {Radice}},\ }\href@noop {}
  {\bibfield  {journal} {\bibinfo  {journal} {Monthly Notices of the Royal
  Astronomical Society}\ }\textbf {\bibinfo {volume} {500}},\ \bibinfo {pages}
  {696} (\bibinfo {year} {2021})}\BibitemShut {NoStop}%
\bibitem [{\citenamefont {Vartanyan}\ \emph {et~al.}(2019)\citenamefont
  {Vartanyan}, \citenamefont {Burrows}, \citenamefont {Radice}, \citenamefont
  {Skinner},\ and\ \citenamefont {Dolence}}]{vartanyan2019successful}%
  \BibitemOpen
  \bibfield  {author} {\bibinfo {author} {\bibfnamefont {D.}~\bibnamefont
  {Vartanyan}}, \bibinfo {author} {\bibfnamefont {A.}~\bibnamefont {Burrows}},
  \bibinfo {author} {\bibfnamefont {D.}~\bibnamefont {Radice}}, \bibinfo
  {author} {\bibfnamefont {M.~A.}\ \bibnamefont {Skinner}}, \ and\ \bibinfo
  {author} {\bibfnamefont {J.}~\bibnamefont {Dolence}},\ }\href@noop {}
  {\bibfield  {journal} {\bibinfo  {journal} {Monthly Notices of the Royal
  Astronomical Society}\ }\textbf {\bibinfo {volume} {482}},\ \bibinfo {pages}
  {351} (\bibinfo {year} {2019})}\BibitemShut {NoStop}%
\bibitem [{\citenamefont {{Lentz}}\ \emph {et~al.}(2015)\citenamefont
  {{Lentz}}, \citenamefont {{Bruenn}}, \citenamefont {{Hix}}, \citenamefont
  {{Mezzacappa}}, \citenamefont {{Messer}}, \citenamefont {{Endeve}},
  \citenamefont {{Blondin}}, \citenamefont {{Harris}}, \citenamefont
  {{Marronetti}},\ and\ \citenamefont {{Yakunin}}}]{Lentz:2015ApJ}%
  \BibitemOpen
  \bibfield  {author} {\bibinfo {author} {\bibfnamefont {E.~J.}\ \bibnamefont
  {{Lentz}}}, \bibinfo {author} {\bibfnamefont {S.~W.}\ \bibnamefont
  {{Bruenn}}}, \bibinfo {author} {\bibfnamefont {W.~R.}\ \bibnamefont {{Hix}}},
  \bibinfo {author} {\bibfnamefont {A.}~\bibnamefont {{Mezzacappa}}}, \bibinfo
  {author} {\bibfnamefont {O.~E.~B.}\ \bibnamefont {{Messer}}}, \bibinfo
  {author} {\bibfnamefont {E.}~\bibnamefont {{Endeve}}}, \bibinfo {author}
  {\bibfnamefont {J.~M.}\ \bibnamefont {{Blondin}}}, \bibinfo {author}
  {\bibfnamefont {J.~A.}\ \bibnamefont {{Harris}}}, \bibinfo {author}
  {\bibfnamefont {P.}~\bibnamefont {{Marronetti}}}, \ and\ \bibinfo {author}
  {\bibfnamefont {K.~N.}\ \bibnamefont {{Yakunin}}},\ }\href {\doibase
  10.1088/2041-8205/807/2/L31} {\bibfield  {journal} {\bibinfo  {journal} {The
  Astrophysical Journal Letters}\ }\textbf {\bibinfo {volume} {807}},\ \bibinfo
  {eid} {L31} (\bibinfo {year} {2015})},\ \Eprint
  {http://arxiv.org/abs/1505.05110} {arXiv:1505.05110 [astro-ph.SR]}
  \BibitemShut {NoStop}%
\bibitem [{\citenamefont {Bollig}\ \emph {et~al.}(2021)\citenamefont {Bollig},
  \citenamefont {Yadav}, \citenamefont {Kresse}, \citenamefont {Janka},
  \citenamefont {M\"uller},\ and\ \citenamefont {Heger}}]{Bollig:2020phc}%
  \BibitemOpen
  \bibfield  {author} {\bibinfo {author} {\bibfnamefont {R.}~\bibnamefont
  {Bollig}}, \bibinfo {author} {\bibfnamefont {N.}~\bibnamefont {Yadav}},
  \bibinfo {author} {\bibfnamefont {D.}~\bibnamefont {Kresse}}, \bibinfo
  {author} {\bibfnamefont {H.~T.}\ \bibnamefont {Janka}}, \bibinfo {author}
  {\bibfnamefont {B.}~\bibnamefont {M\"uller}}, \ and\ \bibinfo {author}
  {\bibfnamefont {A.}~\bibnamefont {Heger}},\ }\href {\doibase
  10.3847/1538-4357/abf82e} {\bibfield  {journal} {\bibinfo  {journal}
  {Astrophys. J.}\ }\textbf {\bibinfo {volume} {915}},\ \bibinfo {pages} {28}
  (\bibinfo {year} {2021})},\ \Eprint {http://arxiv.org/abs/2010.10506}
  {arXiv:2010.10506 [astro-ph.HE]} \BibitemShut {NoStop}%
\bibitem [{\citenamefont {Raffelt}\ and\ \citenamefont
  {Sigl}(1993)}]{Raffelt1993}%
  \BibitemOpen
  \bibfield  {author} {\bibinfo {author} {\bibfnamefont {G.}~\bibnamefont
  {Raffelt}}\ and\ \bibinfo {author} {\bibfnamefont {G.}~\bibnamefont {Sigl}},\
  }\href@noop {} {\bibfield  {journal} {\bibinfo  {journal} {Astroparticle
  Physics}\ }\textbf {\bibinfo {volume} {1}},\ \bibinfo {pages} {165 }
  (\bibinfo {year} {1993})}\BibitemShut {NoStop}%
\bibitem [{\citenamefont {Sigl}\ and\ \citenamefont
  {Raffelt}(1993)}]{Sigl1993}%
  \BibitemOpen
  \bibfield  {author} {\bibinfo {author} {\bibfnamefont {G.}~\bibnamefont
  {Sigl}}\ and\ \bibinfo {author} {\bibfnamefont {G.}~\bibnamefont {Raffelt}},\
  }\href@noop {} {\bibfield  {journal} {\bibinfo  {journal} {Nuclear Physics
  B}\ }\textbf {\bibinfo {volume} {406}},\ \bibinfo {pages} {423 } (\bibinfo
  {year} {1993})}\BibitemShut {NoStop}%
\bibitem [{\citenamefont {Duan}\ \emph {et~al.}(2006)\citenamefont {Duan},
  \citenamefont {Fuller}, \citenamefont {Carlson},\ and\ \citenamefont
  {Qian}}]{duan2006simulation}%
  \BibitemOpen
  \bibfield  {author} {\bibinfo {author} {\bibfnamefont {H.}~\bibnamefont
  {Duan}}, \bibinfo {author} {\bibfnamefont {G.~M.}\ \bibnamefont {Fuller}},
  \bibinfo {author} {\bibfnamefont {J.}~\bibnamefont {Carlson}}, \ and\
  \bibinfo {author} {\bibfnamefont {Y.-Z.}\ \bibnamefont {Qian}},\ }\href@noop
  {} {\bibfield  {journal} {\bibinfo  {journal} {Physical Review D}\ }\textbf
  {\bibinfo {volume} {74}},\ \bibinfo {pages} {105014} (\bibinfo {year}
  {2006})}\BibitemShut {NoStop}%
\bibitem [{\citenamefont {Duan}\ \emph
  {et~al.}(2010{\natexlab{b}})\citenamefont {Duan}, \citenamefont {Fuller},\
  and\ \citenamefont {Qian}}]{duan2010collective}%
  \BibitemOpen
  \bibfield  {author} {\bibinfo {author} {\bibfnamefont {H.}~\bibnamefont
  {Duan}}, \bibinfo {author} {\bibfnamefont {G.~M.}\ \bibnamefont {Fuller}}, \
  and\ \bibinfo {author} {\bibfnamefont {Y.-Z.}\ \bibnamefont {Qian}},\
  }\href@noop {} {\bibfield  {journal} {\bibinfo  {journal} {Annual Review of
  Nuclear and Particle Science}\ }\textbf {\bibinfo {volume} {60}},\ \bibinfo
  {pages} {569} (\bibinfo {year} {2010}{\natexlab{b}})}\BibitemShut {NoStop}%
\bibitem [{\citenamefont {Wolfenstein}(1978)}]{wolfenstein1978neutrino}%
  \BibitemOpen
  \bibfield  {author} {\bibinfo {author} {\bibfnamefont {L.}~\bibnamefont
  {Wolfenstein}},\ }\href@noop {} {\bibfield  {journal} {\bibinfo  {journal}
  {Physical Review D}\ }\textbf {\bibinfo {volume} {17}},\ \bibinfo {pages}
  {2369} (\bibinfo {year} {1978})}\BibitemShut {NoStop}%
\bibitem [{\citenamefont {Mikheev}\ and\ \citenamefont
  {Smirnov}(1985)}]{mikheev1985resonance}%
  \BibitemOpen
  \bibfield  {author} {\bibinfo {author} {\bibfnamefont {S.}~\bibnamefont
  {Mikheev}}\ and\ \bibinfo {author} {\bibfnamefont {A.~Y.}\ \bibnamefont
  {Smirnov}},\ }\href@noop {} {\bibfield  {journal} {\bibinfo  {journal}
  {Yadernaya Fizika}\ }\textbf {\bibinfo {volume} {42}},\ \bibinfo {pages}
  {1441} (\bibinfo {year} {1985})}\BibitemShut {NoStop}%
\bibitem [{\citenamefont {Fano}(1961)}]{fano1961transmission}%
  \BibitemOpen
  \bibfield  {author} {\bibinfo {author} {\bibfnamefont {R.~M.}\ \bibnamefont
  {Fano}},\ }\href@noop {} {\bibfield  {journal} {\bibinfo  {journal} {American
  Journal of Physics}\ }\textbf {\bibinfo {volume} {29}},\ \bibinfo {pages}
  {793} (\bibinfo {year} {1961})}\BibitemShut {NoStop}%
\bibitem [{\citenamefont {Abarbanel}(2013)}]{abarbanel2013predicting}%
  \BibitemOpen
  \bibfield  {author} {\bibinfo {author} {\bibfnamefont {H.}~\bibnamefont
  {Abarbanel}},\ }\href@noop {} {\emph {\bibinfo {title} {Predicting the
  future: completing models of observed complex systems}}}\ (\bibinfo
  {publisher} {Springer},\ \bibinfo {year} {2013})\BibitemShut {NoStop}%
\bibitem [{\citenamefont {Oden}\ and\ \citenamefont
  {Reddy}(2012)}]{oden2012variational}%
  \BibitemOpen
  \bibfield  {author} {\bibinfo {author} {\bibfnamefont {J.~T.}\ \bibnamefont
  {Oden}}\ and\ \bibinfo {author} {\bibfnamefont {J.~N.}\ \bibnamefont
  {Reddy}},\ }\href@noop {} {\emph {\bibinfo {title} {Variational methods in
  theoretical mechanics}}}\ (\bibinfo  {publisher} {Springer Science \&
  Business Media},\ \bibinfo {year} {2012})\BibitemShut {NoStop}%
\bibitem [{\citenamefont {Ye}\ \emph {et~al.}(2015)\citenamefont {Ye},
  \citenamefont {Rey}, \citenamefont {Kadakia}, \citenamefont {Eldridge},
  \citenamefont {Morone}, \citenamefont {Rozdeba}, \citenamefont {Abarbanel},\
  and\ \citenamefont {Quinn}}]{ye2015systematic}%
  \BibitemOpen
  \bibfield  {author} {\bibinfo {author} {\bibfnamefont {J.}~\bibnamefont
  {Ye}}, \bibinfo {author} {\bibfnamefont {D.}~\bibnamefont {Rey}}, \bibinfo
  {author} {\bibfnamefont {N.}~\bibnamefont {Kadakia}}, \bibinfo {author}
  {\bibfnamefont {M.}~\bibnamefont {Eldridge}}, \bibinfo {author}
  {\bibfnamefont {U.~I.}\ \bibnamefont {Morone}}, \bibinfo {author}
  {\bibfnamefont {P.}~\bibnamefont {Rozdeba}}, \bibinfo {author} {\bibfnamefont
  {H.~D.}\ \bibnamefont {Abarbanel}}, \ and\ \bibinfo {author} {\bibfnamefont
  {J.~C.}\ \bibnamefont {Quinn}},\ }\href@noop {} {\bibfield  {journal}
  {\bibinfo  {journal} {Physical Review E}\ }\textbf {\bibinfo {volume} {92}},\
  \bibinfo {pages} {052901} (\bibinfo {year} {2015})}\BibitemShut {NoStop}%
\bibitem [{\citenamefont {W{\"a}chter}(2009)}]{wachter2009short}%
  \BibitemOpen
  \bibfield  {author} {\bibinfo {author} {\bibfnamefont {A.}~\bibnamefont
  {W{\"a}chter}},\ }in\ \href@noop {} {\emph {\bibinfo {booktitle} {Dagstuhl
  Seminar Proceedings}}}\ (\bibinfo {organization} {Schloss
  Dagstuhl-Leibniz-Zentrum f{\"u}r Informatik},\ \bibinfo {year}
  {2009})\BibitemShut {NoStop}%
\bibitem [{min()}]{minAone}%
  \BibitemOpen
  \href@noop {} {\enquote {\bibinfo {title} {{minAone interface with
  Interior-point Optimizer}},}\ }\bibinfo {howpublished}
  {\url{https://github.com/yejingxin/minAone}},\ \bibinfo {note} {accessed:
  2020-05-26}\BibitemShut {NoStop}%
\bibitem [{\citenamefont {AA-Ahmetaj}()}]{github}%
  \BibitemOpen
  \bibfield  {author} {\bibinfo {author} {\bibnamefont {AA-Ahmetaj}},\
  }\href@noop {} {\enquote {\bibinfo {title} {Slurm\_minaone},}\ }\bibinfo
  {howpublished}
  {\url{https://github.com/AA-Ahmetaj/SLURM_minAone}}\BibitemShut {NoStop}%
\end{thebibliography}%


\end{document}